# Hidden low-temperature magnetic order revealed through magnetotransport in monolayer CrSBr

*Evan J. Telford[1,2,†], Avalon H. Dismukes[1,†], Raymond L. Dudley[2], Ren A. Wiscons[1], Kihong Lee[1], Jessica Yu[1], Sara Shabani[2], Allen Scheie[3], Kenji Watanabe[4], Takashi Taniguchi[4], Di Xiao[5], Abhay N. Pasupathy[2], Colin Nuckolls[1], Xiaoyang Zhu[1], Cory R. Dean[2]\*, Xavier Roy[1]\**

1 - Department of Chemistry, Columbia University, New York, NY 10027, USA.
2 - Department of Physics, Columbia University, New York, NY 10027, USA.
3 - Neutron Scattering Division, Oak Ridge National Laboratory, Oak Ridge, Tennessee 37831, USA.
4 - National Institute for Materials Science, 1-1 Namiki, Tsukuba, 305-0044 Japan.
5 - Department of Physics, Carnegie Mellon University, Pittsburgh, PA 15213, USA.
† - these authors contributed equally

**Abstract:**

**Magnetic semiconductors are a powerful platform for understanding, utilizing and tuning the interplay between magnetic order and electronic transport. Compared to bulk crystals, two-dimensional magnetic semiconductors have greater tunability, as illustrated by the gate modulation of magnetism in exfoliated $CrI_3$ and $Cr_2Ge_2Te_6$, but their electrically insulating properties limit their utility in devices. Here we report the simultaneous electrostatic and magnetic control of electronic transport in atomically-thin CrSBr, an A-type antiferromagnetic semiconductor. Through magnetotransport measurements, we find that spin-flip scattering from the interlayer antiferromagnetic configuration of multilayer flakes results in giant negative magnetoresistance. Conversely, magnetoresistance of the ferromagnetic monolayer CrSBr vanishes below the Curie temperature. A second transition ascribed to the ferromagnetic ordering of magnetic defects manifests in a large positive magnetoresistance in the monolayer and a sudden increase of the bulk magnetic susceptibility. We demonstrate this magnetoresistance is tunable with an electrostatic gate, revealing that the ferromagnetic coupling of defects is carrier mediated.**

**Main Text:**

Layered A-type antiferromagnets are composed of van der Waals (vdW) sheets with intralayer ferromagnetic (FM) order and interlayer antiferromagnetic (AFM) coupling[1]. Upon the application of an external magnetic field, the interlayer AFM order can be switched to a field-induced FM configuration, accompanied by a change in optical and electronic properties[2–6]. This change of spin structure produces emergent phenomena, including giant tunneling magnetoresistance in vertical vdW spin-filters[4,7,8], giant second harmonic generation (SHG) in the AFM state due to the breaking of inversion symmetry by magnetic order[9], and magnetic order-dependent excitonic transitions arising from changes in interlayer hybridization[2]. For applications in spin-based electronics, ideal materials should combine layered magnetism with functional semiconducting transport properties, which would allow for simultaneous control over charge and spin carriers. In studies of bulk magnetic semiconductors, magnetic defects and impurities play a crucial role in determining the magnetic and electronic properties. To further develop 2D magnetic semiconductors, it is thus critical to understand how magnetic order and magnetic defects couple to charge carriers. Transport measurements in currently available 2D magnets, however, are limited to FM metals[10,11] or degenerately-doped FM semiconductors[12], while the role of defects is essentially unexplored.

In this work, we report the magnetotransport properties of atomically-thin CrSBr, a 2D vdW material that combines layered A-type AFM order and semiconducting transport properties. Each CrSBr layer consists of two buckled rectangular planes of CrS fused together, with both surfaces capped by Br atoms (**Figure 1A**)[3,13,14]. Stacking of the layers along the *c*-axis produces an orthorhombic structure with space group *Pmmn*. The paramagnetic (PM)-to-AFM phase transition includes significant intralayer FM correlations above the Néel temperature ($T_N$ = 132 K) developing at the bulk Curie temperature (160 K), as identified by heat capacity[9] and magnetic susceptibility[3] measurements. Below $T_N$, the layers order ferromagnetically along the *b*-axis and align antiferromagnetically along the stacking direction (**Figure 1A**)[3,13]. Crystals of CrSBr can be mechanically exfoliated and a recent SHG study confirms that the bulk magnetic structure persists down to the FM monolayer and AFM bilayer[9].

In bulk single crystals, CrSBr is an extrinsic semiconductor with a direct bandgap of ~1.5 eV and finite conductivity that can be measured down to liquid helium temperature[3]. We find the transport properties of few-layer CrSBr is dominated by the interlayer magnetic order. When the flakes are polarized with an external magnetic field, their resistances decrease drastically due to differences in interlayer spin-flip scattering between the AFM and FM configurations. In monolayer CrSBr, spin-flip scattering arises only from intraplanar FM ordering, which manifests as a peak of negative magnetoresistance near the monolayer Curie temperature ($T_C$ = 146 K[9]), followed by a drop to near zero upon cooling to ~40 K. For all layer numbers, magnetoresistance measurements reveal an unexpected magnetic phase below 40 K, which we identify as carrier-mediated FM ordering of magnetic defects. In monolayer CrSBr we controllably switch between magnetoresistance mechanisms attributed to s-d exchange interactions and bound magnetic polarons by varying the carrier density with an electrostatic gate.

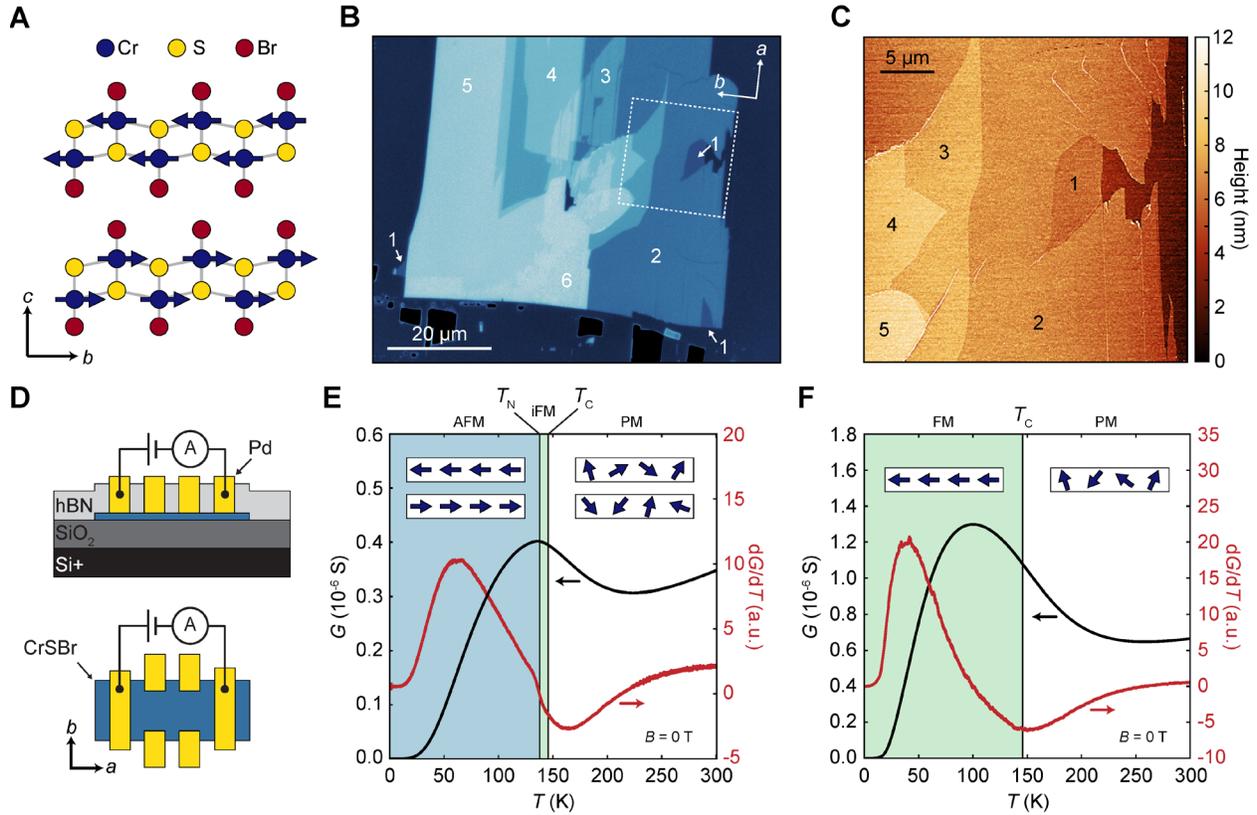

**Figure 1. Crystal structure, device fabrication, and transport signatures of CrSBr magnetic ordering. A)** Crystal structure of CrSBr as viewed along the *a*-axis. Orientation of the Cr spins in the AFM state are given by solid blue arrows. Blue, yellow, and red circles correspond to Cr, S, and Br, respectively. **B)** False-colored optical image of an exfoliated CrSBr flake with thicknesses ranging from 1 to 6 layers. The corresponding layer numbers are denoted on the image. The orientation of the crystal axes is given in the upper right inset. **C)** Atomic force microscopy image of the CrSBr flake shown in (**B**). The region where the image was taken is denoted by a dashed white box in (**B**). The corresponding number of CrSBr layers is labelled on the plot. **D)** Side-view (top) and top-view (bottom) schematic of the CrSBr device geometry. Color code: hBN, grey; Pd, yellow; $SiO_2$, dark grey; CrSBr, blue; Si+ substrate, black. The orientation of the crystal axes relative to the electrodes is denoted. **E, F)** Conductance (solid black line) and derivative of the conductance (solid red line) versus *T* at zero *B* for bilayer (**E**) and monolayer (**F**) CrSBr. The interlayer AFM, intralayer FM (iFM/FM), and PM states are denoted by solid blue, solid green, and white regions, respectively. $T_N$ is defined as the location of the kink in $dG/dT$. $T_C$ is defined from SHG measurements[9]. Cartoons of the spin orientation in each state are given in the insets. The white rectangles represent single CrSBr sheets, and the blue arrows represent the Cr spins.

Atomically-thin CrSBr flakes are prepared via mechanical exfoliation on Si wafers with a 285 nm thick $SiO_2$ layer (see Supplemental for details)[15,16]. The thickness and crystallographic directions of exfoliated flakes are determined by optical contrast (**Figure 1B** and **Supplemental**

**Figures 1-3**), atomic force microscopy (**Figure 1C** and **Supplemental Figure 3**), and Raman spectroscopy (**Supplemental Figures 4,5**). Mesoscopic transport devices are fabricated using the *via* contact method[17], whereby palladium electrodes embedded in hexagonal boron nitride (hBN) are transferred onto the desired CrSBr flakes using the dry-polymer-transfer process[18] (**Figure 1D**; see Supplemental for details). We performed electrical transport measurements as a function of temperature ($T$), magnetic field ($B$), and electrostatic gate voltage ($V_{BG}$) on CrSBr flakes ranging in thickness from 1 to 9 layers. Current was sourced along the crystallographic *a*-axis for all measurements (**Figure 1D**). Owing to the high resistance of the flakes over the entire $T$ range (**Supplemental Figures 6,7**), all data reported in the main text were measured in a 2-terminal configuration. Some measurements were repeated in a 4-terminal configuration to confirm the channel resistance dominates the transport properties (**Supplemental Figure 8**).

Bilayer CrSBr displays an overall extrinsic semiconducting behavior with conductance ($G$) decreasing with decreasing $T$ (**Figure 1E**). At $T = 136 \pm 4$ K, there is a sharp kink in the derivative of $G$ versus $T$ (d$G$/d$T$), which is attributed to the onset of AFM ordering at $T_N$[3]. $G$ shows a local maximum at the same $T$ due to reduced scattering caused by spin fluctuations as CrSBr becomes antiferromagnetically ordered[19,20]. Thicker flakes display the same kink in d$G$/d$T$ versus $T$, signaling the onset of AFM order (**Supplementary Figure 9**). Within experimental error, the values of $T_N$ measured from transport for flakes ranging in thickness from 2 to 9 layers are unchanged from the bulk value and independent of layer number (**Supplemental Figure 9**). In contrast, monolayer CrSBr shows no kink in d$G$/d$T$ (**Figure 1F**), but displays a local minimum close to the expected monolayer $T_C$[9]. This is consistent with previous reports that monolayer CrSBr only exhibits intraplanar FM ordering[9]. Note that the low-$T$ resistance for both bilayer and monolayer CrSBr is well-described by Efros-Shklovskii variable range hopping, indicating electronic transport is dominated by electron percolation between hopping sites (**Supplemental Figure 10**)[21,22].

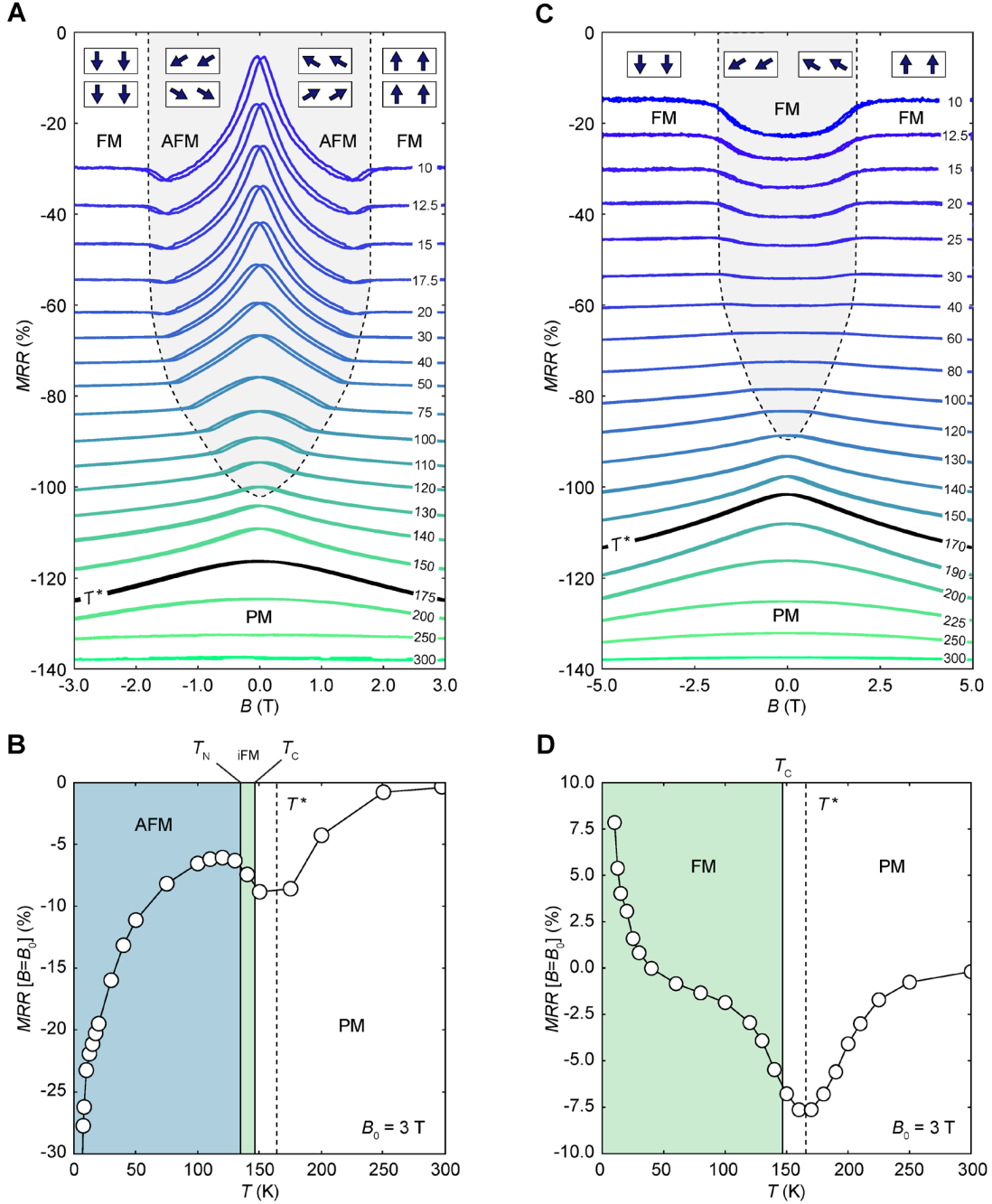

**Figure 2. Magnetoresistance of bilayer and monolayer CrSBr. A, C)** *MRR* versus *B* at various *T* with *B* oriented along the *c*-axis for bilayer (**A**) and monolayer (**C**) CrSBr. Both forward and backward *B* sweeps are presented. The curves are offset for clarity. The solid black line is the curve taken near *T**, the temperature at which *MRR* has a local minimum. The corresponding magnetic phases are labeled, and $B_{sat}$ is denoted by a dashed black line. Insets: schematics showing the orientation of the spins in each state. The white rectangles represent single CrSBr sheets, and the blue arrows represent the Cr spins. **B, D)** *MRR* at a fixed *B* parallel to the *c*-axis versus *T* for bilayer (**B**) and monolayer (**D**) CrSBr. The *B* at which the fixed-field *MRR* is calculated is $B_0 = 3$ T. The interlayer AFM, intralayer FM (iFM/FM), and PM phases are labelled and denoted by blue, green, and white regions, respectively.

**Figure 2A** presents the magnetoresistance ratio (*MRR*) versus *B* and *T* for bilayer CrSBr. We define $MRR = \frac{R(B)-R(B=0)}{R(B=0)} \times 100$, where in this plot *B* is oriented along the *c*-axis. From 300 to ~175 K, the sample is in a PM phase characterized by a broad negative *MRR* (n*MRR*) due to the field-induced suppression of spin-flip scattering between conducting electrons and local magnetic moments[23,24]. The thermal fluctuations which prevent spins from aligning with *B* diminish with decreasing *T*, leading to an overall increase in the magnitude of the n*MRR*. Below $T_N = 136 \pm 4$ K, we observe n*MRR* up to a well-defined saturation field ($B_{sat}$) beyond which the device resistance saturates with increasing *B*. This manifests as a dome of n*MRR*, the edges of which define $B_{sat}$ (dashed black line in **Figure 2A**). The magnitude of n*MRR* increases with decreasing *T* as the AFM state becomes more ordered, reaching –23.5% at 10 K (**Figure 2B**; see **Supplemental Figure 11** for lower *T*). This giant n*MRR* followed by saturation for $B > B_{sat}$ indicates that carrier scattering between layers is controlled by the interlayer magnetic configuration. In the AFM state at zero *B*, interlayer tunneling is suppressed. As *B* increases, the spins are gradually canted towards the *c*-axis, progressively breaking the AFM configuration and restoring interlayer tunneling, which results in a decrease of the device resistance. At $B_{sat}$, the spins are fully polarized along the c-axis, and a further increase of *B* has little effect on the resistance. Consistent with this understanding, all CrSBr samples thicker than 1 layer exhibit the same qualitative *MRR* behavior (see **Supplemental Figures 12-15**).

**Figure 2C** plots the magnetoresistance of the FM CrSBr monolayer versus *B* and *T*. Between 300 and 170 K, monolayer CrSBr exhibits the same behavior as multilayer samples; n*MRR* which we attribute to the suppression of spin-flip scattering in the PM state with increasing *B*, the magnitude of which increases as *T* decreases. As *T* is further lowered below 170 K, the magnitude of *MRR* decreases and approaches zero by 40 K (**Figure 2D**). This is expected since the intralayer FM order reduces spin fluctuations at zero *B*, diminishing spin-flip scattering. The competition between these two phenomena leads to a peak in n*MRR* versus *T* at 170 K (black line in **Figure 2C**), which we denote as *T**. This feature is present in all samples (black line in **Figure 2A** for bilayer and bulk[3]), reflecting the onset of intraplanar FM correlations as it closely follows the monolayer $T_C$ measured by SHG and bulk heat capacity[9]. *T** is the same for bilayer and monolayer CrSBr, indicating that this feature is independent of the interlayer AFM coupling. The remarkable agreement between transport and previous SHG results demonstrates that magnetotransport is a reliable probe of both FM and AFM order in CrSBr. Below 40 K, monolayer

CrSBr exhibits a positive *MRR* (p*MRR*) that increases with decreasing *T* (**Figure 2D**). While this feature is most prominent in monolayer samples, it is also observed in the bilayer CrSBr sample in **Figure 2A** (and in bulk CrSBr[3]) as a small p*MRR* just below $B_{sat}$ (see **Supplemental Figure 16** for a detailed analysis).

This low-*T* p*MRR* response is unexpected for a FM monolayer whose ordering temperature is well above 100 K. To better understand the origin of the p*MRR*, we performed magnetometry on bulk single crystals of CrSBr and complementary field-angle-dependent transport measurements on monolayer CrSBr (**Figure 3**). The bulk magnetic susceptibility ($\chi$) versus *T* curves (**Figure 3A**) show the expected cusp at 134 ± 2 K associated with the AFM transition, followed by a sharp increase of $\chi$ onsetting at ~35 K (defined as $T_D$), which is unusual for an A-type antiferromagnet. The crystal structure of CrSBr is unchanged across this transition[3] and we observe an increase in $\chi$ for all *B* directions, which necessarily exempts structural transformation or spin reorientation as the origin of this feature. We also note that the increase in $\chi$ below $T_D$ resembles a FM transition. Consistent with this hypothesis, there is a significant difference between the zero-field-cooled and field-cooled traces of $\chi$ versus *T* across $T_D$ (inset of **Figure 3A**). In **Figure 3B**, we plot the magnetization (*M*) versus *B* at 2 K. The overall response is dominated by the AFM behavior[3,13], characterized by a spin-flip transition at ~0.3 T along the easy *b*-axis, and gradual canting of spins along the non-easy axes up to $B_{sat}$. However, if we focus on the low-*B* region along the easy axis, we observe the emergence of a small sigmoidal hysteresis below $T_D$, characteristic of FM ordering. The development of an additional FM phase is further evidenced by plotting d*M*/d*B* versus *B* (**Figure 3C**). Above $T_D$ (40 K), d*M*/d*B* is constant for all *B* directions up to ~0.2 T but below $T_D$ (2 K), there is an additional contribution to d*M*/d*B* at low *B*[25].

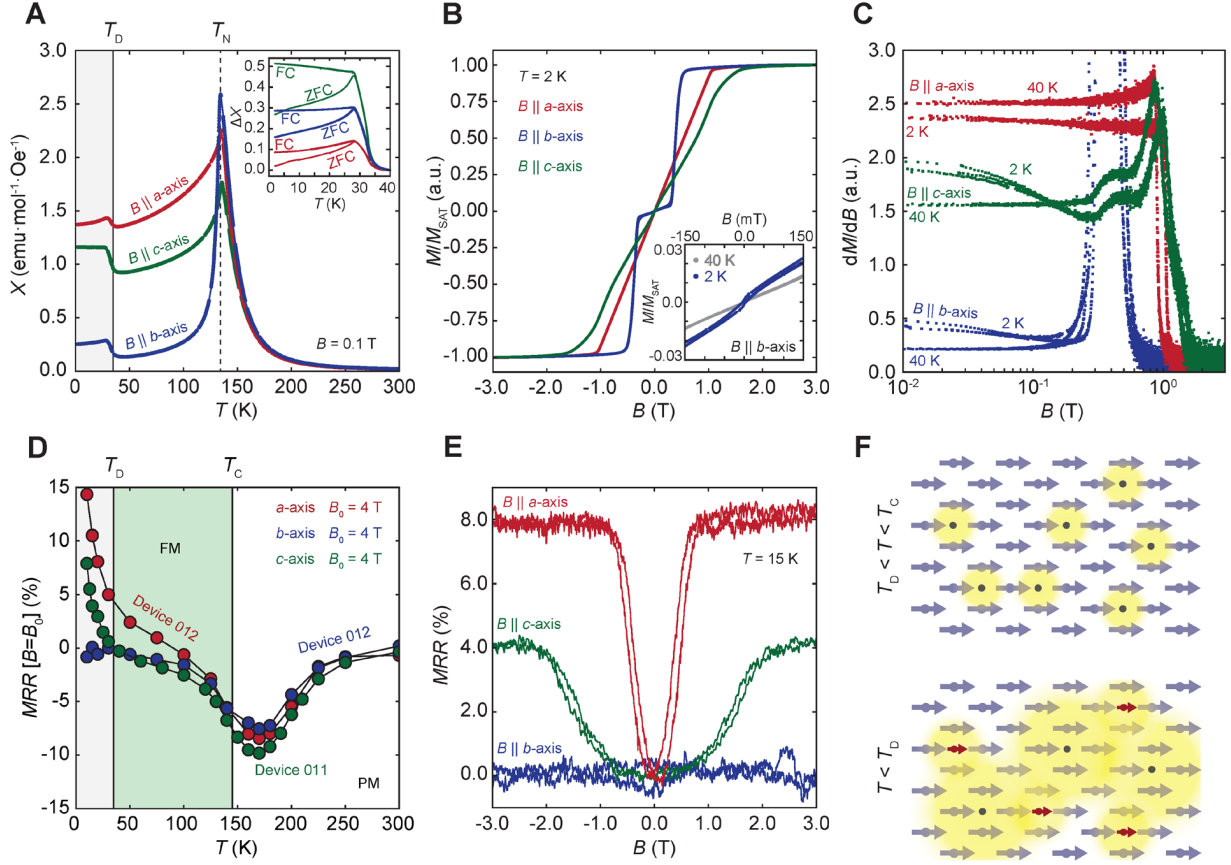

**Figure 3. Evidence for ferromagnetic ordering of magnetic defects in CrSBr. A)** Bulk $\chi$ versus $T$ for $B = 0.1$ T oriented along the $a$-axis (red), $b$-axis (blue), and $c$-axis (green). Top right inset: plot of the change in zero-field-cooled (ZFC) and field-cooled (FC) susceptibility across $T_D$ with $B = 0.01$ T. **B)** Bulk $M$ versus $B$ at 2 K for $B$ oriented along the $a$-axis (red), $b$-axis (blue) and $c$-axis (green). The lower right inset plots the low-$B$ $M$ versus $B$ for $B$ along the $b$-axis above (40 K – grey dots) and below (2 K – blue dots) $T_D$. **C)** d$M$/d$B$ versus $B$ on a log scale above (40 K) and below (2 K) $T_D$ for $B$ oriented along the $a$-axis (red), $b$-axis (blue), and $c$-axis (green). **D)** MRR at a fixed $B$ versus $T$ for $B$ oriented along the $a$-axis (red dots), $b$-axis (blue dots), and $c$-axis (green dots). The $B$ at which the fixed-field MRR is calculated for each $B$ direction is given in the legend. The FM, PM, and ordered defect phase are labelled by green, white, and grey regions, respectively. **E)** MRR versus $B$ in monolayer CrSBr for $B$ oriented along the $a$-axis (red), $b$-axis (blue), and $c$-axis (green). All curves were taken at 15 K and both forward and backward $B$ sweeps are presented. **F)** Schematic of the spin structure of monolayer CrSBr for $T_D < T < T_C$ (top) and $T < T_D$ (bottom). Cr spins, defect sites, and polarized magnetic defect spins are denoted as blue arrows, black dots, and red arrows, respectively. The yellow clouds represent the localization radius of the carriers.

The *MRR* of monolayer CrSBr depends on the direction of $B$ below $T_D$ (**Figure 3E**). The p*MRR* response is characterized by a quadratic dependence at low $B$ followed by a saturation at $B_{sat}$ when $B$ is oriented along the intermediate axis ($a$-axis) and the hard axis ($c$-axis). When $B$ is along the easy axis ($b$-axis), there is negligible *MRR*, indicating that the spins are aligned along the $b$-axis at zero $B$. The p*MRR* along the $a$- and $c$-axes therefore arises from a canting of the spins from the $b$-axis towards the respective field directions, supported by the observation that the *MRR* saturation fields closely match the saturation fields in the bulk $M$ versus $B$ curves (**Figure 3B**).

The difference in magnitude of p*MRR* between the *a*- and *c*-axes is likely due to the anisotropic magnetoresistance (AMR) effect, which produces p*MRR* when the magnetization direction is parallel to the source current direction[26]. For 40 K < *T* < 100 K, the data is consistent with the AMR effect; *MRR* > 0 for fields parallel to the *a*-axis and *MRR* ~0 for fields parallel to the *c*- and *b*-axes (**Figure 3D**). Below 40 K, the *MRR* along the *a*- and *c*-axes increases drastically (**Figure 3D**). The same anisotropy in *MRR* was previously observed in bulk single-crystal transport measurements, in which p*MRR* features emerge along the *a*- and *c*-axes below ~40 K, but its origin was not explained[3].

The coincidence of the FM phase with the onset of p*MRR* is consistent with carrier-mediated FM ordering of magnetic defects[27,28]. In analogous bulk magnetic semiconductor systems, magnetic defects can order collectively as a result of exchange interactions with localized charge carriers[29–31] or with the intrinsic magnetic lattice[32,33]. In CrSBr, the magnetic structure likely consists of the dominant Cr magnetic lattice and a sublattice of coupled defect spins. The magnetometry data indicates that upon cooling below $T_N$, Cr spins within each layer order ferromagnetically, but the defects remain unpolarized (**Figure 3F: top**). At $T_D$, the magnetic defects become polarized and adopt the same FM configuration as the Cr lattice due to the strong exchange interaction between the Cr spins and magnetic defects (**Figure 3F: bottom**). The fact that $T_D$ is much lower than $T_N$ strongly suggests that the defect moments (red arrows in **Figure 3F**) arise from self-trapped electrons near donor sites (known as magnetic polarons[32,33]) although we cannot entirely exclude the possibility of intrinsic magnetic point defects.

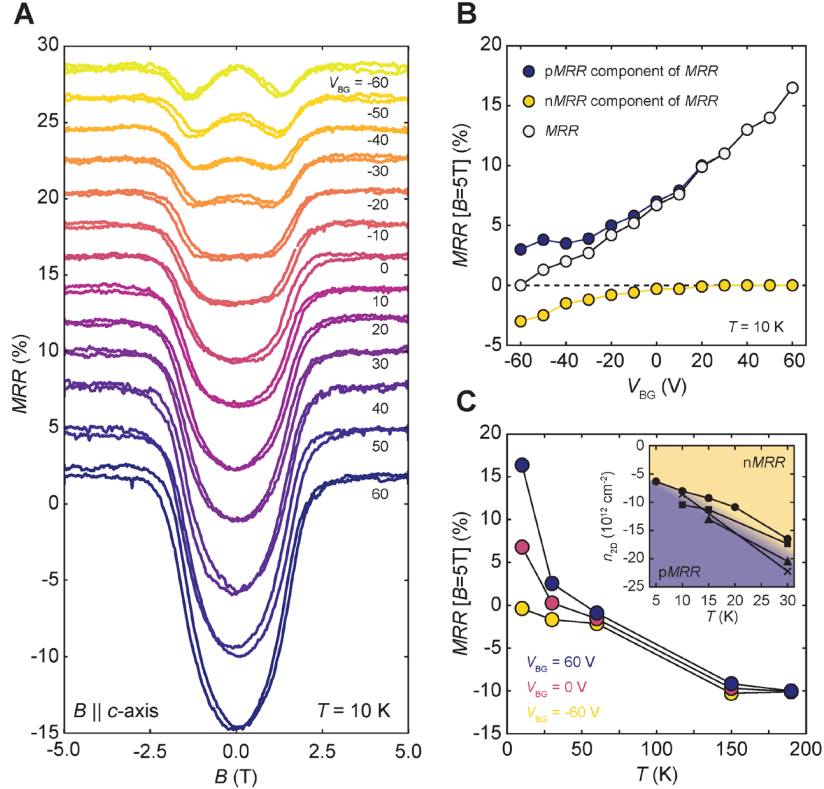

**Figure 4. Electrostatic control of magnetoresistance in monolayer CrSBr. A)** Monolayer CrSBr *MRR* versus *B* oriented along the *c*-axis for various $V_{BG}$ at 10 K. Both forward and backward *B* sweeps at each $V_{BG}$ are presented. The curves are offset for clarity. The value of $V_{BG}$ for each trace is denoted. **B)** *MRR* at 5 T (grey dots) and the corresponding extracted p*MRR* (blue dots) and n*MRR* (yellow dots) components versus $V_{BG}$ at 10 K. **C)** *MRR* at 5 T versus *T* for $V_{BG}$ = 60 V (blue dots), $V_{BG}$ = 0 V (pink dots), and $V_{BG}$ = –60 V (yellow dots). A phase diagram depicting the crossover between p*MRR* (blue region) and n*MRR* (yellow region) versus *T* and intrinsic sheet carrier density ($n_{2D}$) is given in the inset. Each data marker type corresponds to a different monolayer CrSBr device (4 in total - black squares, black crosses, black circles, and black triangles).

The magnetic defect coupling strength and the corresponding *MRR* response is predicted to strongly depend upon carrier density, which in our device can be dynamically and reversibly tuned using an electrostatic gate. **Figure 4A** presents the *MRR* of monolayer CrSBr versus *B* at different $V_{BG}$. Starting from 0 V, as we increase $V_{BG}$ (increase electron density), we observe a significant increase in the p*MRR* response, from 7.6% up to 16.4%. Conversely, decreasing $V_{BG}$ (decreasing electron density) decreases the p*MRR*, and a noticeable region of n*MRR* emerges below –20 V at low *B*. At a gate voltage of –60 V, the *MRR* reaches –0.4% with regions of n*MRR* followed by p*MRR* equal in magnitude when $B < B_{sat}$. **Figure 4B** plots the *MRR* versus $V_{BG}$ along with the extracted p*MRR* and n*MRR* contributions. There is a direct competition between the n*MRR* and p*MRR* regimes; increasing the electron density ($V_{BG} > 0$) yields larger p*MRR*, while decreasing the electron density ($V_{BG} < 0$) diminishes the p*MRR* contribution and induces n*MRR*. Together, these produce a *MRR* that depends linearly on $V_{BG}$. The *T* dependence of the *MRR* at different $V_{BG}$ (**Figure 4C** and **Supplemental Figure 18**) shows that the slope of *MRR* versus $V_{BG}$ decreases quickly with increasing *T* and is negligible above $T_D$. The sensitivity to doping is further supported by the air-sensitivity of monolayer CrSBr flakes. By exposing a monolayer device to air for a period of three weeks (**Supplemental Figure 19**), we were able to change the doping level and the sign and magnitude of the *MRR* at low *T* and reproduce the behavior of a second monolayer device with lower intrinsic carrier density (**Supplemental Figures 20-23**). The carrier density at which the *MRR* crosses from p*MRR* to n*MRR* increases as *T* increases and is consistent between multiple samples (inset of **Figure 4C**).

The sensitivity of the *MRR* to carrier density supports the hypothesis that the magnetotransport properties of CrSBr at low *T* are governed by the carrier-mediated FM phase. The two competing magnetoresistance features (p*MRR* and n*MRR*) indicates that the unique shape of the *MRR* curves results from a competition between multiple mechanisms that dominate within certain carrier density ranges. The n*MRR* component is characteristic of magnetic polarons[27,28,34–36] and is consistent with our observation that decreasing the carrier density enhances the n*MRR*[37]. Additionally, the carrier density at which *MRR* crosses from p*MRR* to n*MRR* (inset of **Figure 4C**) is close to the estimated magnetic defect density (see Supplemental for calculation), implying that we observe n*MRR* when the carrier density is less than the magnetic defect density[29]. However, the formation of magnetic polarons is generally not associated with a large p*MRR*. In dilute magnetic semiconductors, p*MRR* often arises from a *B*-dependent broadening of the defect density of states due to s-d exchange between magnetic impurities and separate electronic defect states[27,28,35,36,38,39]. The coexistence of these two *MRR* mechanisms suggests that CrSBr contains both magnetic and non-magnetic defects. Both s-d exchange and magnetic polarons are mechanisms consistent with the emergence of a FM phase and electron transport in the variable range hopping limit[27,28,34,38].

Magnetotransport is typically an indirect probe of magnetism but in CrSBr it reveals in striking detail a rich magnetic structure and its intricate and tunable coupling to charge carriers. For bilayer and thicker flakes, $T_N$ is independent of layer number and giant n*MRR* emerges as the interlayer AFM configuration is broken. In monolayer CrSBr, intraplanar FM order gives rise to a peak of n*MRR* close to $T_C$. This n*MRR* feature is also present in multilayer flakes, signaling the existence of intraplanar FM correlations above $T_N$. We uncovered a hidden magnetic phase below 40 K ascribed to the FM ordering of magnetic defects. In monolayer CrSBr, this ordered magnetic defect phase dominates the low-$T$ magnetoresistance, which can be linearly tuned using an electrostatic gate. These results highlight the utility of CrSBr as a 2D magnet with giant intrinsic n$MRR$. In the monolayer, the sensitivity of $MRR$ to carrier density presents a unique opportunity not only for fabricating tunable spintronic devices, but also for understanding and utilizing defects to engineer tunable properties in vdW magnets.

**Methods:**
See Supplemental Materials

**Corresponding Authors:**
*Xavier Roy (e-mail: xr2114@columbia.edu)
*Cory R. Dean (e-mail: cd2478@columbia.edu)

**Acknowledgements:**
We would like to thank Jordan Pack for his help in using the double-axis rotator for the field-direction-dependent transport measurements. We would also like to thank Jiang Xiao for helpful discussions in interpreting our transport data. Research on magnetotransport properties of vdW magnetic semiconductors was supported as part of Programmable Quantum Materials, an Energy Frontier Research Center funded by the U.S. Department of Energy (DOE), Office of Science, Basic Energy Sciences (BES), under award DE-SC0019443. A portion of this research used resources at the SNS, a Department of Energy (DOE) Office of Science User Facility operated by ORNL. A.H.D. was supported by the NSF graduate research fellowship program (DGE 16-44869). R.A.W. was supported by the Arnold O. Beckman Fellowship in Chemical Sciences. The Columbia University Shared Materials Characterization Laboratory (SMCL) was used extensively for this research. We are grateful to Columbia University for the support of this facility. The PPMS used to perform vibrating sample magnetometry and some of the transport measurements was purchased with financial support from the NSF through a supplement to award DMR-1751949.

**Author Contributions:**

EJT and AHD prepared the CrSBr flakes. EJT and AHD performed the optical contrast calibration and atomic force microscopy measurements. EJT and RLD fabricated the transport devices and performed the transport measurements. AHD synthesized the bulk crystals. EJT and KL performed the Raman spectroscopy measurements. EJT, AHD, and RAW performed the vibrating sample magnetometry measurements. EJT performed the oxidation measurements. All authors contributed to analyzing the data and writing the manuscript.

**Competing Interests:**

The authors declare no competing financial interest.

**Supporting Information:**

**Hidden low-temperature magnetic order revealed through magnetotransport in monolayer CrSBr**

*Evan J. Telford[1,2,†], Avalon H. Dismukes[1,†], Raymond L. Dudley[2], Ren A. Wiscons[1], Kihong Lee[1], Jessica Yu[1], Sara Shabani[2], Allen Scheie[3], Kenji Watanabe[4], Takashi Taniguchi[4], Di Xiao[5], Abhay N. Pasupathy[2], Colin Nuckolls[1], Xiaoyang Zhu[1], Cory R. Dean[2]\*, Xavier Roy[1]\**

1 - Department of Chemistry, Columbia University, New York, NY 10027, USA.
2 - Department of Physics, Columbia University, New York, NY 10027, USA.
3 - Neutron Scattering Division, Oak Ridge National Laboratory, Oak Ridge, Tennessee 37831, USA.
4 - National Institute for Materials Science, 1-1 Namiki, Tsukuba, 305-0044 Japan.
5 - Department of Physics, Carnegie Mellon University, Pittsburgh, PA 15213, USA.
† - these authors contributed equally

**Exfoliation and Flake Identification**

CrSBr flakes were exfoliated onto 285 nm or 90 nm $SiO_2/Si+$ substrates using mechanical exfoliation with Scotch® Magic™ tape[1,2]. For > 1 L devices, $SiO_2/Si+$ substrates were exposed to a gentle oxygen plasma for 5 minutes to remove adsorbates from the surface and increase flake adhesion[3]. The exfoliation was done under ambient conditions by heating the mother tape for 3 minutes at 100º C, letting it cool to room temperature, then peeling the tape from the substrate as quickly as possible[3]. For 1 L devices, the $SiO_2/Si+$ substrates were passivated by depositing a thin layer of 1-dodecanol before exfoliation[4]. The exfoliation was done under inert conditions in an $N_2$ glovebox with < 1 ppm $O_2$ and < 1 ppm $H_2O$ content. The mother tape was placed onto the $SiO_2/Si+$ substrates without heating and removed as quickly as possible. CrSBr flake thickness was identified using optical contrast before encapsulation and then confirmed with atomic force microscopy after encapsulation with hexagonal boron nitride (hBN).

**Optical Contrast Calbration**

To more quickly and reliably identify the thickness of CrSBr flakes, a contrast calibration curve was developed for both 285 nm and 90 nm $SiO_2/Si+$ substrates. First, a series of images was collected of various CrSBr flakes with varying thicknesses using a Nikon Eclipse LV150N microscope and Nikon DS-Fi3 camera. The images were then shading corrected in which the inhomogeneous illumination of the substrate across a single image was corrected by dividing an optical image of a pristine area of the chip without CrSBr flakes. The contrast of the flakes was then extracted using Gwyddion to measure the difference in RGB color between the substrate and the desired flake. We found that the red color contrast was the most significant, so all reported optical contrasts are with respect to red. The series of extracted contrasts were binned into a histogram and the histrogram was fitted to an N-peak gaussian, where N is the number of expected flake thicknesses. The extracted positions of the gaussian peaks is the average red optical contrast for each CrSBr thickness (**Supplemental Figures 1,2**). The thicknesses of the flakes were confirmed with atomic force microscopy (**Supplemental Figure 3**).

**Atomic Force Microscopy**

Atomic force microscopy was performed in a Bruker Dimension Icon® using OTESPA-R3 tips in tapping mode. Flake thicknesses were extracted using Gwyddion to measure histograms of the height difference between the substrate and the desired CrSBr flake.

**Raman Spectroscopy**

*> 1 L CrSBr*
Raman spectroscopy for CrSBr flakes > 1 L was performed under ambient conditions in a Renishaw InVia™ micro-Raman microscope using a 532 nm wavelength laser. A 50x objective was used with a laser spot size of 2-3 µm . A laser power of 100 µW was used with a grating of 2400 g/mm for all spectra. Varying acquisition times were used depending on the flake thickness (longer times for thinner flakes). For each flake, 10 spectra were acquired and averaged after subtracting a dark background. The dark background was a spectra acquired with no laser excitation and the same acquisition parameters.

*1 L CrSBr*

Raman spectroscopy for 1 L CrSBr flakes was performed inside an $N_2$ glovebox with < 5 ppm $O_2$ and < 0.5 ppm $H_2O$ with a Horiba XploRA™ Raman microscope using a 532 nm wavelength laser. A x100 objective was used with a laser spot size of ~1-2 μm. A laser power of ~20 μW was used with a grating of 2400 g/mm for all spectra. For each flake, 5 spectra were acquired with an acquisition time of 180 s and averaged after subtracting a dark background. The dark background was a spectra acquired with no laser excitation and the same acquisition parameters.

**Transport Device Fabrication**

Transport devices were fabricated from CrSBr flakes using the via contact technique[5] in which hBN with embedded palladium electrodes was placed onto the desired CrSBr flake using the dry-polymer-transfer technique[6]. For > 1 L CrSBr flakes, the transfer process was performed under ambient conditions. For monolayer CrSBr flakes, the transfer process was performed under inert conditions in an $N_2$ glovebox with < 5 ppm $O_2$ and < 0.5 ppm $H_2O$. Bonding pads were then designed and deposited using conventional electron beam lithography and deposition techniques. All devices were diced by hand and bonded to a 16-pin DIP socket for measurement in cryogenic systems. Between fabrication steps, 1 L devices were stored in the $N_2$ glovebox to avoid sample degredation.

**Electrical Transport Measurements**

Longitudinal resistance was measured in a 2-terminal configuration using an SRS830 lock-in amplifier to source voltage and measure current using a 17.777 Hz reference frequency. Four-terminal longitudinal and Hall measurements were performed in a four-terminal configuration using SRS830 lock-in amplifiers to source voltage, measure current, and measure the Hall voltage using a 17.777 Hz reference frequency. Due to the morphology of the exfoliated crystals, the current and longitudinal resistances were measured parallel to the *a*-axis and the Hall resistance was measured parallel to the *b*-axis. Variable temperatures between 1.6 K and 300 K and magnetic fields between -9 T and 9 T were achieved in a Janis pumped $^4$He cryostat. Sample temperature equilibrium was checked by monitoring sample resistivity for stability over time at a fixed temperature. Hysteresis in the superconducting magnet due to trapped fields was measured by identifying the zero-field shift from the forward and backward field scans measured in the non-magnetic state (at $T = 300$ K). This hysteresis was accounted for in all presented magnetoresistance measurements. For measurements without electrostatic gating, the silicon back gate was kept grounded using a grounding cap. For gate-dependent measurements, a Keithley 2400 was used to output voltages between -60 V and 60 V on the silicon back gate. A protection resistor of 100 kOhm was placed in series between the gate and the voltage source.

The transmission line measurements (TLMs) were performed at room temperature by sourcing voltage and measuring current with a Keithley 2400. A voltage excitation of 0.5 V was used to ensure measurements were performed in the linear IV regime.

The field-direction-dependent transport measurements on monolayer CrSBr were performed using a home-made double-axis rotator in the aforementioned Janis pumped $^4$He cryostat. The angle was adjusted at room temperature inside the cryostat to avoid exposing the sample to air between measurement runs. Once the angle was set at room temperature, it was kept fixed for all transport measurements along the chosen crystallographic axis. The angle of the magnetic field relative to the crystallographic axes of CrSBr was determined by orienting the sample relative to the mounted DIP socket by hand and aligning the DIP socket to the probe axis by eye. We estimate an uncertainty in the field angle of ~3°.

The field-direction-dependent transport measurement on bulk CrSBr (**Supplemental Figure 24**) were performed in a Quantum Design PPMS® DynaCool™ system using the electrical transport option coupled with a single-axis rotator probe. To align the magnetic field to all three crystal axes, two devices fabricated from the same single CrSBr crystal were mounted in orthogonal orientations on a single rotator puck. In this configuration, the field was aligned to the *c*-axis for both samples at 0°. At 90°, the field was aligned to *b*-axis for one device and the *a*-axis for the second device. All transport data was collected in a 2-terminal geometry using an AC excitation between 0.1 and 2 V and an AC frequency <50 Hz. IV curves were performed at each measurement temperature to ensure a linear response. Due to the morphology of the crystals, current was driven along the *a*-axis for all measurements. The uncertainty in the field angle is ~1°.

**Vibrating Sample Magnetometry**

All vibrating sample magnetometry (VSM) was conducted on a Quantum Design PPMS® DynaCool™ system. A single CrSBr crystal was selected and the surface was exfoliated mechanically to expose a pristine interface. The crystal was attached to a quartz paddle using GE varnish (which was cured at room temperature under ambient conditions for 30 minutes) and oriented with the *a*-, *b*-, or *c*- axis perpendicular to the length of the quartz paddle. The same crystal was used for all axial orientated measurements. The variable temperature scans and field-dependent magnetic susceptibility curves for each axis were measured during the same measurement cycle. The crystal was removed using a 1:1 ethanol/toluene solution, dried in air, then reoriented and reattached using the previously prescribed varnish method. Each full-range variable temperature scan was programmed as follows using the DynaCool™ VSM module: 1) demagnetization of the SC magnet at 300 K by sweeping the field from 20000 Oe to 0 Oe with an oscillatory field ramp, 2) magnetic field set to 1000 Oe using a linear field ramp, 3) cooled to 2 K at 12 K/min, 3) measured susceptibility versus temperature upon warming with a ramp rate of 5 K/min. The field dependent-magnetic susceptibility curves at different temperatures were programmed as follows using the DynaCool™ VSM module: 1) demagnetization of the SC magnet at 300 K by sweeping the field from 20000 Oe to 0 Oe with an oscillatory field ramp, 2) cooled to the desired temperature at 12 K/min, 3) measured magnetization versus field from -50000 Oe to 50000 Oe over 3 cycles (0 to -50000, -50000 to 50000, 50000 to -50000, -50000 to 0). The low-temperature zero-field-cooled and field-cooled variable temperature scans were programmed as follows using the DynaCool™ VSM module: 1) demagnetization of the SC magnet at 300 K by sweeping the field from 20000 Oe to 0 Oe with an oscillatory field ramp, 2) cooled to 2 K at 12 K/min, 3) set the magnetic field to 100 Oe, 4) measure susceptibility versus temperature from 2 K up to 40 K ramping the temperature at 1 K/min, 5) measure susceptibility versus temperature from 40 K down to 2 K ramping the temperature at 1 K/min, 6) re-measure susceptibility versus temperature again from 2 K up to 40 K ramping the temperature at 1 K/min.

The density of magnetic impurities was estimated by converting the absolute change in susceptibility (in units of emu/mol/Oe) across $T_D$ assuming all impurities undergo ferromagnetic ordering and have a spin of ½:

$$n_D = \frac{\Delta\chi * B}{\mu_B} * \frac{N_M}{V}$$

Where $\Delta\chi$ is the change in susceptibility, $\mu_B$ is the Bohr magneton (in CGS units), $B$ is the applied magnetic field, $V$ is the crystal volume, and $N_M$ is the number of moles of CrSBr. The estimated defect density is ~$10^{13}$ cm$^{-2}$.

**Calculating Carrier Density from Gate Dependence**

The carrier density at low temperatures was estimated using the gate-dependence of the sample conductivity. Assuming the conductivity varies linearly with carrier density, we use the capacitor model to estimate the intrinsic doping density.

$$\sigma \propto n$$

$$n = n_D + \frac{\epsilon_0 \epsilon_R}{d} V_G$$

Where $n_D$ is the intrinsic doping density, $\varepsilon_0$ is the vacuum permitivity, $\varepsilon_r$ is the relative permitivity of the dielectric (3.9 for SiO$_2$), $d$ is the thickness of the dielectric, and $V_G$ is the value of the back-gate voltage in volts. Assuming the density is the only term that varies with gate voltage in the expression for conductance, an expression for the density can be derived from the derivative of conductance versus gate.

$$\frac{d\sigma}{dV_G} \cdot \frac{1}{\sigma(V_G)} = \frac{dn}{dV_G} \cdot \frac{1}{n(V_G)}$$

Using the expression for density versus gate, we can derive an expression for the intrinsic doping density.

$$n_D = \frac{1}{e} \cdot \left[ \frac{1}{\frac{d\sigma}{dV_G} \cdot \frac{1}{\sigma(V_G)}} - V_G \right] \cdot \frac{\epsilon_0 \epsilon_r}{d}$$

Here, $e$ is the electron charge, $d\sigma/dV_G$ is the measured slope of the conductance versus back-gate voltage curve, and $\sigma(V_G)$ is the conductance at a given back-gate voltage. For the estimated densities in **Supplemental Figure 22**, $V_G$ was set to 0.

**Details of Oxidation Dependent Measurement**

The controlled oxidation of an encapsulated CrSBr device was performed by storing the sample under ambient conditions and performing a TLM at regular intervals. For the device in **Supplemental Figure 19**, the measurements were performed at room temperature by sourcing voltage and measuring current with a Keithley 2400. A voltage excitation of 0.5 V was used to ensure measurements were performed in the linear IV regime. The contact resistance and sample resistivity were extracted by fitting the curves of 2-terminal resistance versus channel length to a line following the TLM model.

$$R = \frac{\rho L}{A} + 2R_C$$

Where $R$ is the measured 2-terminal device resistance, $L$ is the channel length, $A$ is the cross sectional channel area, $\rho$ is the sample resistivity, and $R_C$ is the contact resistance. The contact resistance is multiplied by two since the measurements are 2-terminal measurements between two contacts. In this model, $\rho/A$ is simply the slope of the 2-terminal resistance versus channel length and the intercept is twice the contact resistance.

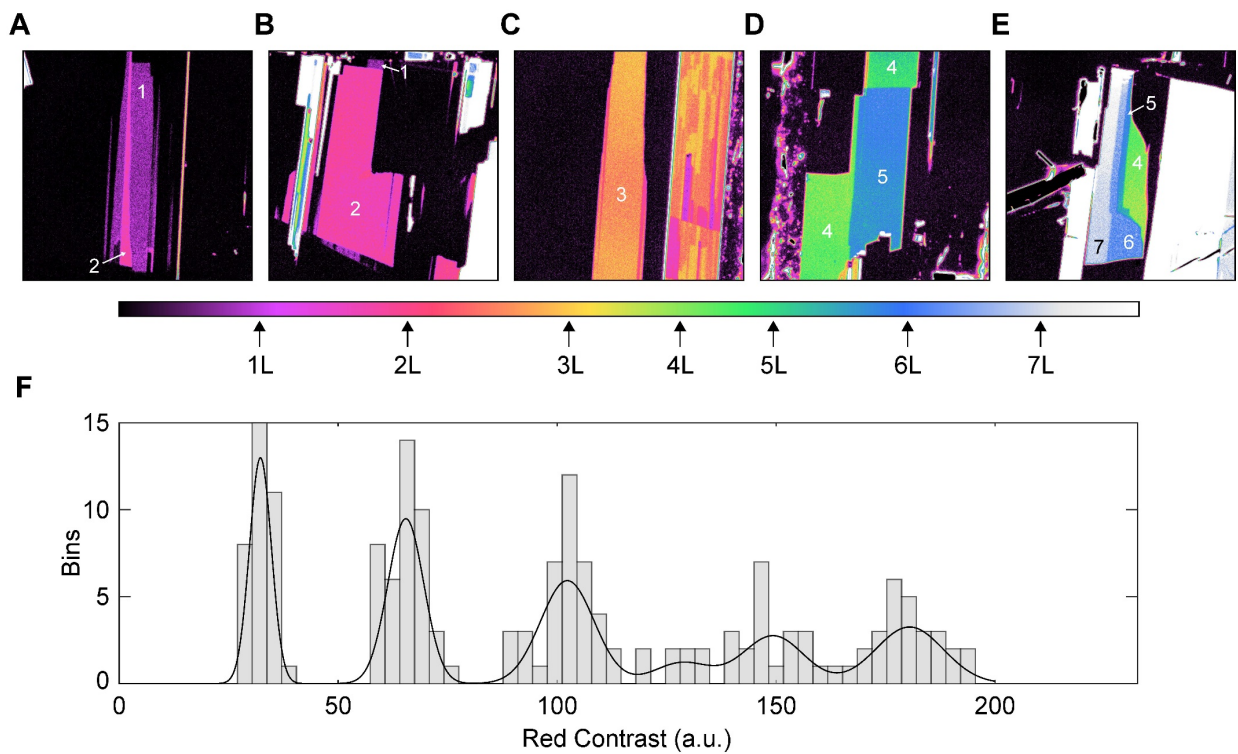

**Figure 1: CrSBr optical contrast calibration on 285 nm SiO$_2$. A-E)** False-colored optical images of various CrSBr flakes ranging in thickness from 1 to 7 layers exfoliated onto 285 nm SiO$_2$. The corresponding layer numbers are labelled on each image. **F)** Histogram of the optical contrast for all cataloged CrSBr flakes < 7 L. The solid black line is a 6 peak Gaussian fit to the data. The color bar above the histogram is a conversion from numerical contrast value to false color in (**A-E**). The colors corresponding to each layer number are denoted by arrows.

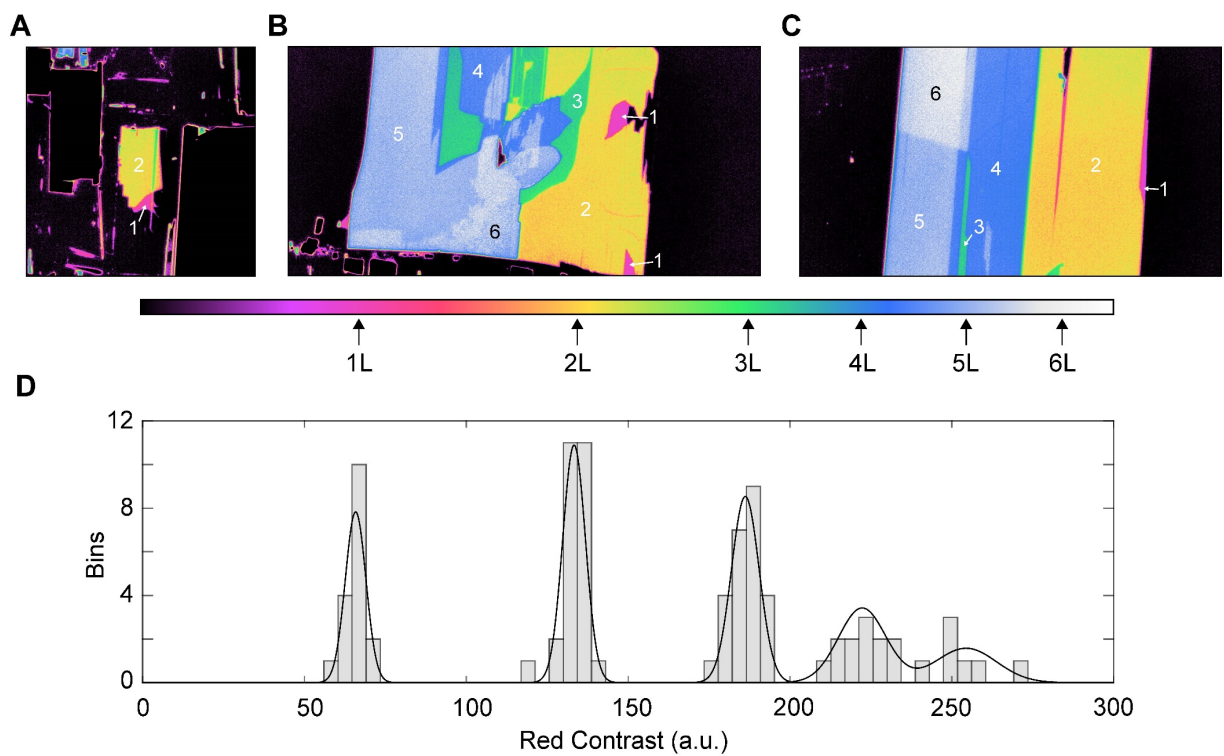

**Figure 2: CrSBr optical contrast calibration on 90 nm SiO$_2$. A-C)** False-colored optical images of various CrSBr flakes ranging in thickness from 1 to 6 layers exfoliated onto 90 nm SiO$_2$. The corresponding layer numbers are labelled on each image. **D)** Histogram of the optical contrast for all cataloged CrSBr flakes < 6 L. The solid black line is a 5 peak Gaussian fit to the data. The color bar above the histogram is a conversion from numerical contrast value to false color in **(A-C)**. The colors corresponding to each layer number are denoted by arrows.

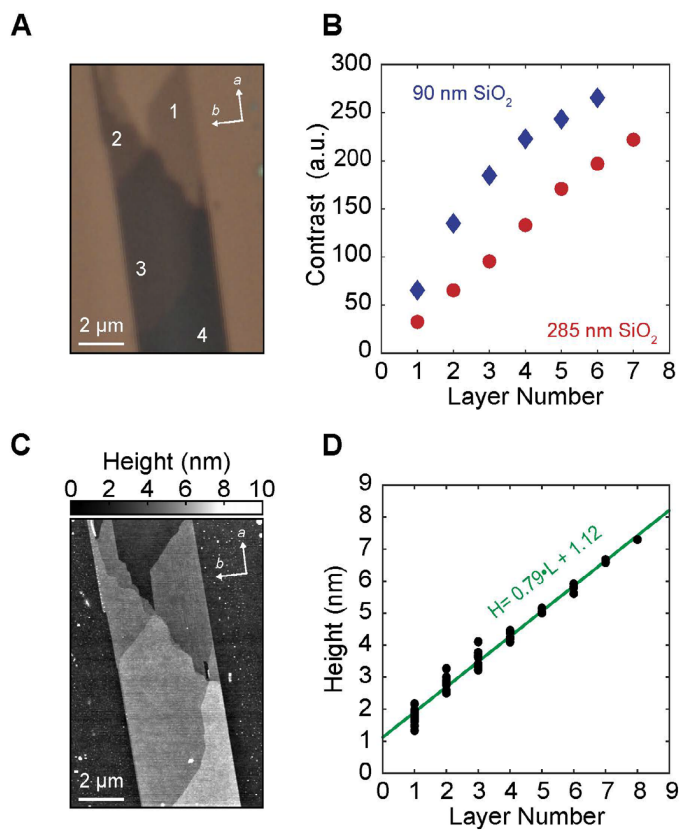

**Figure 3: CrSBr thickness versus layer number correlated to optical contrast. A)** Optical image of a CrSBr flake exfoliated onto 90 nm $SiO_2$. The thickness ranges from 1 to 4 layers with the corresponding layer numbers marked on the image. The orientation of the crystal axes is given in the inset. **B)** Optical contrast versus layer number for CrSBr exfoliated on 285 nm $SiO_2$ (red dots) and 90 nm $SiO_2$ (blue diamonds). **C)** Atomic force microscopy topography of the flake in (**A**). **D)** CrSBr flake thickness versus layer number measured through atomic force microscopy. A linear fit to the data is given by a solid green line with the corresponding fit parameters given in the inset.

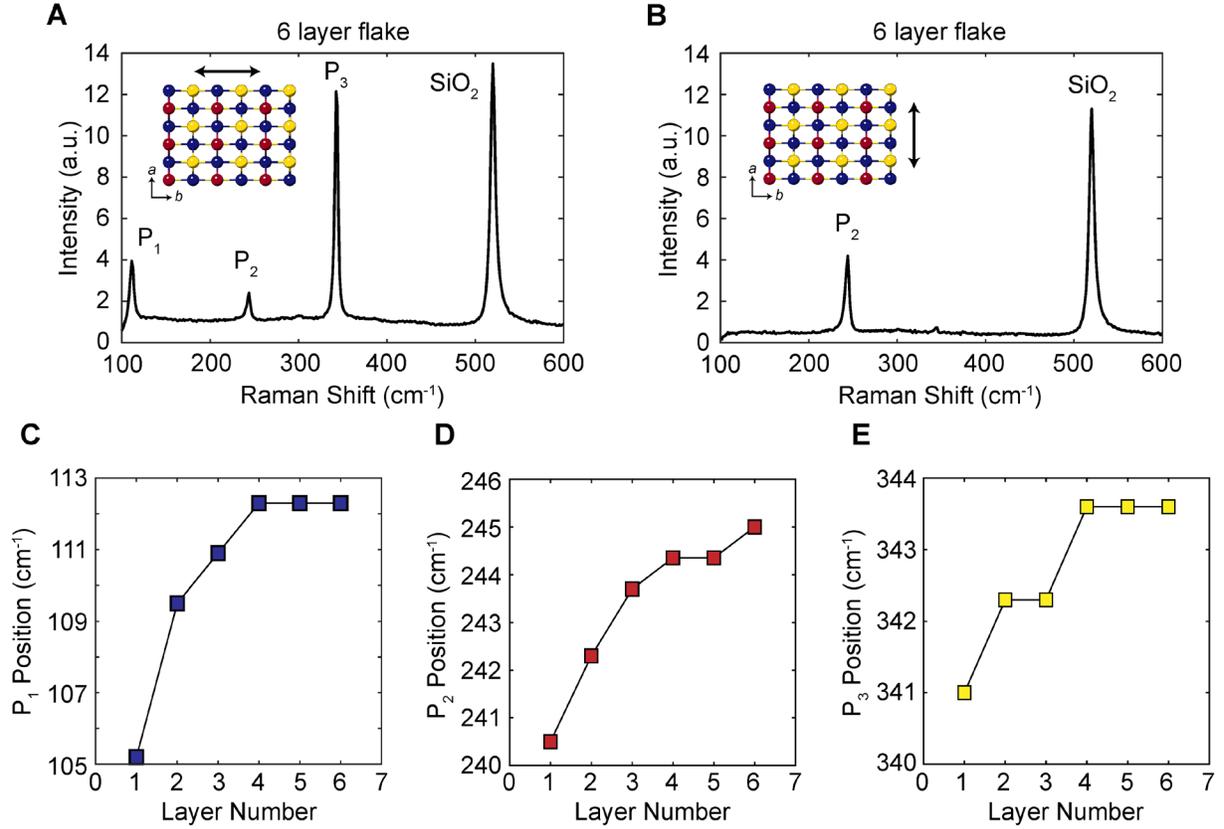

**Figure 4: Layer dependence of CrSBr Raman spectra. A, B)** Raman intensity versus wavenumber for a 6-layer CrSBr flake with incident light polarized parallel to the *b*-axis (**A**) and the *a*-axis (**B**). The Raman peaks from CrSBr and the SiO$_2$ substrate are labelled. **C-E)** Extracted peak positions versus CrSBr thickness for P$_1$ (**C**), P$_2$ (**D**), and P$_3$ (**E**). The peaks are denoted in (**A**) and (**B**).

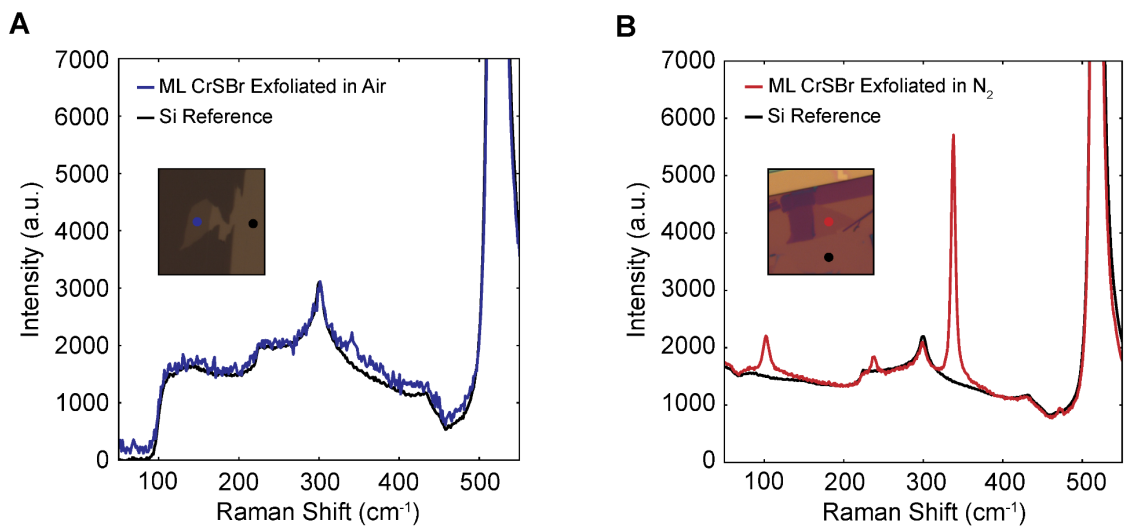

**Figure 5: Raman spectra of pristine and oxidized monolayer CrSBr. A)** Raman intensity versus wavenumber for a monolayer CrSBr flake exfoliated in air (solid blue line) and the $SiO_2$ substrate (solid black line). An optical image of the flake is given in the inset. The positions where the Raman spectra were acquired are labelled by a blue (monolayer CrSBr) and black ($SiO_2$ substrate) dot. **B)** Raman intensity versus wavenumber for a monolayer CrSBr flake exfoliated inside a $N_2$ glovebox (solid red line) and the $SiO_2$ substrate (solid black line). An optical image of the flake is given in the inset. The positions where the Raman spectra were acquired are labelled by a red (monolayer CrSBr) and black ($SiO_2$ substrate) dot.

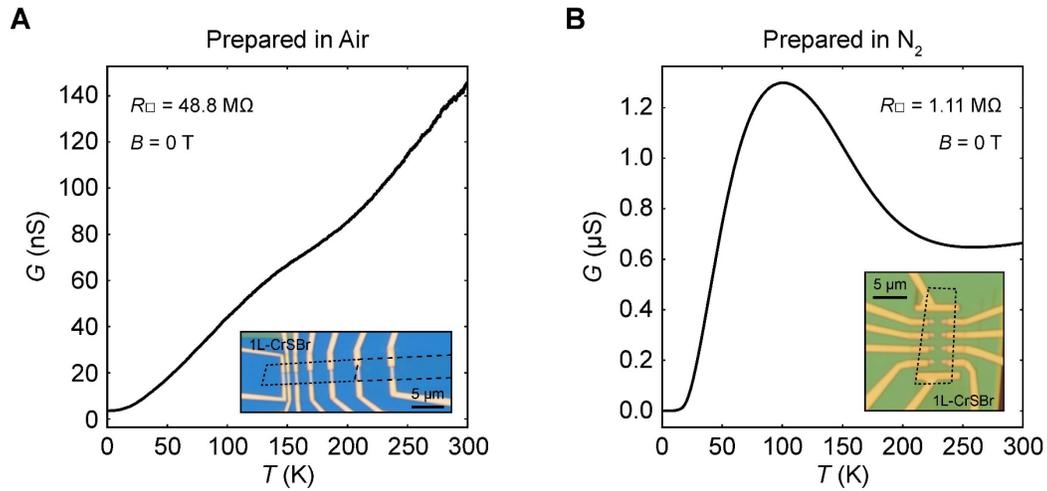

**Figure 6: Transport properties of pristine and oxidized monolayer CrSBr. A)** Conductance versus temperature for a monolayer CrSBr flake exfoliated and encapsulated under ambient conditions. The room temperature sheet resistivity and an optical image of the device is given in the inset. **B)** Conductance versus temperature for the monolayer CrSBr flake presented in the main text. It was exfoliated and encapsulated inside a $N_2$ glovebox on a passivated substrate. The room temperature sheet resistivity and an optical image of the device is given in the inset.

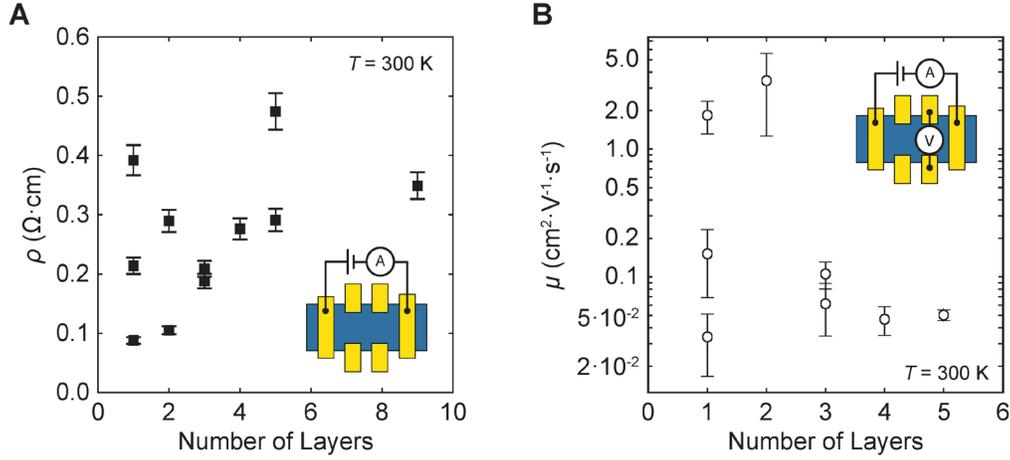

**Figure 7: Layer dependence of CrSBr resistivity and mobility at room temperature. A)** Resistivity of CrSBr flakes versus layer number. The resistivity was estimated by measuring the 2-probe sample resistance across the longest channel length to minimize the size of the contact resistance relative to the channel resistance. The contact resistance to CrSBr was extracted from a TLM for the 5 L and 1 L devices and confirmed to be a negligible percentage of the measured 2-probe resistances (assuming the contact resistance for 1 L CrSBr is an upper bound on the contact resistance for > 1 layers). The error bar denotes the estimated contribution of the contact resistance to the calculated resistivity. The physical dimensions of the flakes were measured with atomic force microscopy. A schematic of the measurement setup is given in the inset. **B)** Calculated sample mobility of CrSBr flakes versus layer number. The mobility was calculated using $\mu = \frac{\sigma}{ne}$. The density was determined from the Hall effect. The large error bars in the 2 L and 1 L samples are due to uncertainties in the Hall measurement from high sample resistance and relatively low sample mobility. A schematic of the measurement setup is given in the inset.

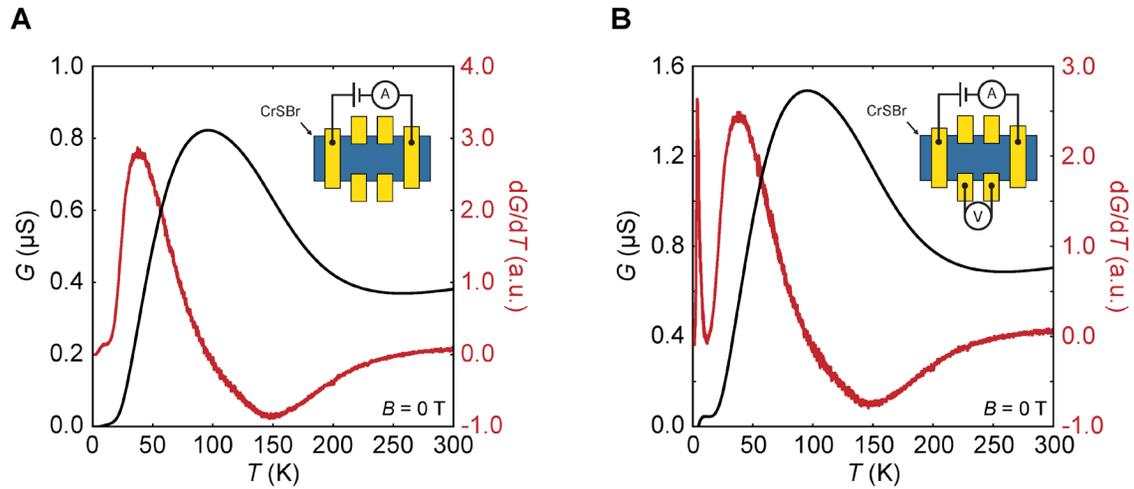

**Figure 8: Comparison of 2-probe and 4-probe zero-field conductance versus temperature.
A, B)** Conductance (solid black line) and derivative of the conductance (solid red line) versus temperature at zero magnetic field for a monolayer CrSBr device measured in a 2-terminal configuration (**A**) and a 4-terminal configuration (**B**). Schematics of the measurement geometries are given in the insets.

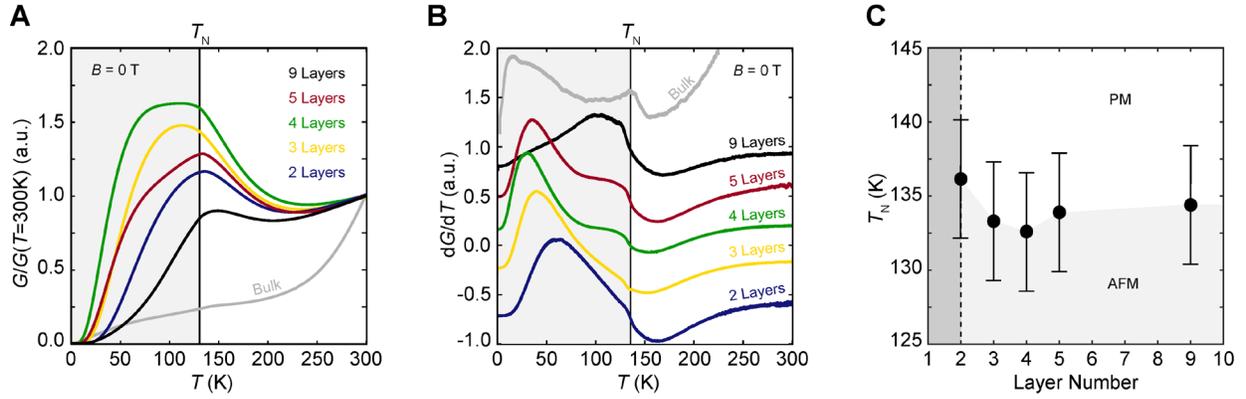

**Figure 9: Layer dependence of CrSBr zero-field conductance versus temperature. A)** Conductance versus temperature for 2 (solid blue line), 3 (solid yellow line), 4 (solid green line), 5 (solid red line), and 9 (solid black line) layers of CrSBr. The bulk data[7] is given as the solid grey line for reference. Each conductance trace is normalized to the value at 300 K. **B)** Derivative of conductance versus temperature for 2 (solid blue line), 3 (solid yellow line), 4 (solid green line), 5 (solid red line), and 9 (solid black line) layers of CrSBr. The bulk data[7] is given as the solid grey line for reference. Each curve is offset from one another for clarity. **C)** Néel temperature versus CrSBr flake thickness. Light grey and white regions correspond to antiferromagnetic and paramagnetic regions, respectively. The dark grey region denotes the ferromagnetic phase observed in the monolayer.

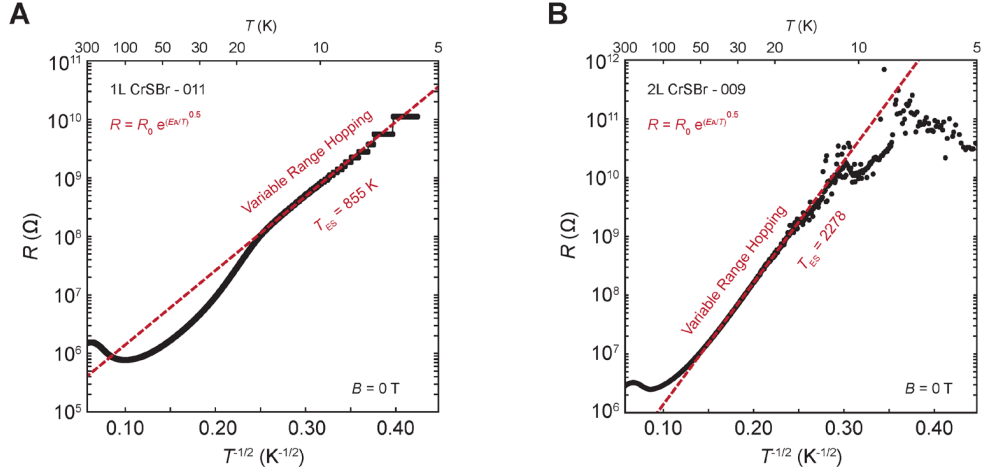

**Figure 10: Analysis of low-temperature resistance behavior in CrSBr. A, B)** Resistance on a log scale versus inverse square root of temperature at zero magnetic field for the monolayer (**A**) and bilayer (**B**) CrSBr presented in the main text. A fit to a variable range hopping model is given by the dashed red line. The extracted Efros-Shklovskii temperature is given in the inset. In both plots, a few absolute temperature values are denoted on the top of each plot.

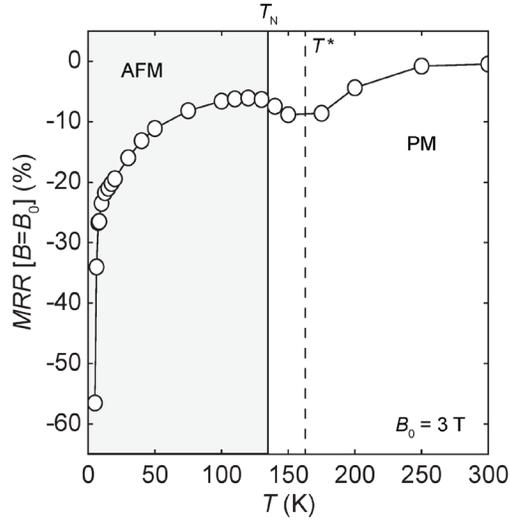

**Figure 11: Bilayer CrSBr fixed-field *MRR* versus temperature.** *MRR* [$B$ = 3 T] versus temperature for the bilayer CrSBr device presented in the main text. The entire data range is shown to emphasize the size of the n*MRR* at the lowest temperatures measured. The AFM and PM phases are denoted by grey and white regions, respectively. The magnetic transitions $T_N$ and $T^*$ are labelled on the plot as a solid and dashed black line, respectively.

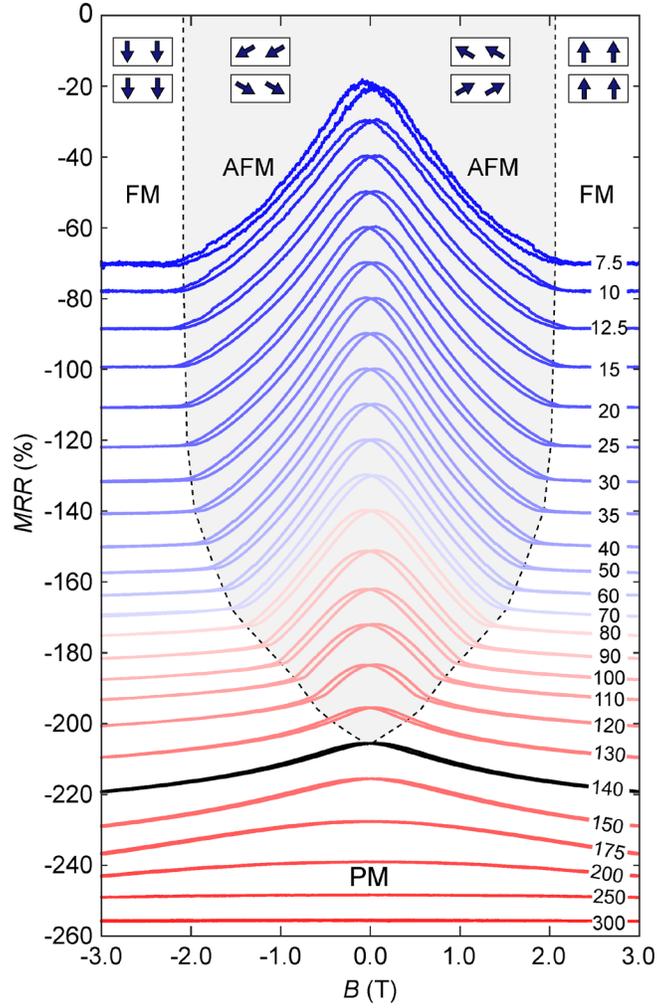

**Figure 12: Magnetotransport properties of 9 L CrSBr.** Magnetoresistance ratio defined as $MRR(B) = \frac{R(B)-R(B=0)}{R(B=0)} * 100$ versus magnetic field at various temperatures with the field oriented along the *c*-axis. Both forward and backward magnetic field scans at each temperature are presented. The curves are offset for clarity. The solid black line is the curve taken at a temperature near $T_N$. The AFM, FM, and PM phases are labeled, and the phase boundary is denoted by a dashed black line. Schematics showing the orientation of the spins in the AFM and FM state are given in the inset.

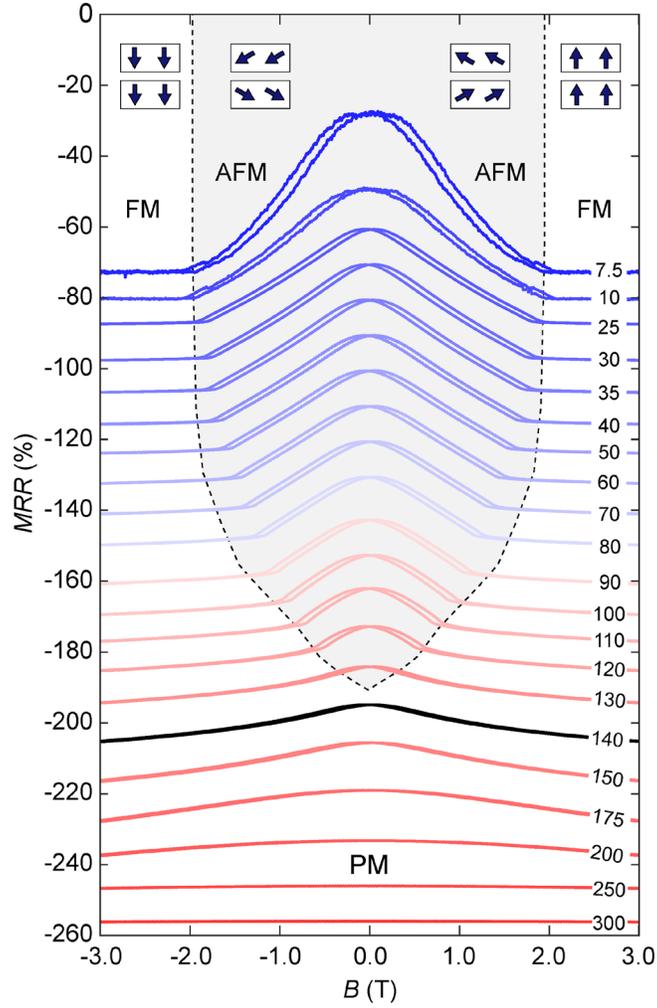

**Figure 13: Magnetotransport properties of 5 L CrSBr.** Magnetoresistance ratio defined as $MRR(B) = \frac{R(B)-R(B=0)}{R(B=0)} * 100$ versus magnetic field at various temperatures with the field oriented along the *c*-axis. Both forward and backward magnetic field scans at each temperature are presented. The curves are offset for clarity. The solid black line is the curve taken at a temperature near $T_N$. The AFM, FM, and PM phases are labeled, and the phase boundary is denoted by a dashed black line. Schematics showing the orientation of the spins in the AFM and FM state are given in the inset.

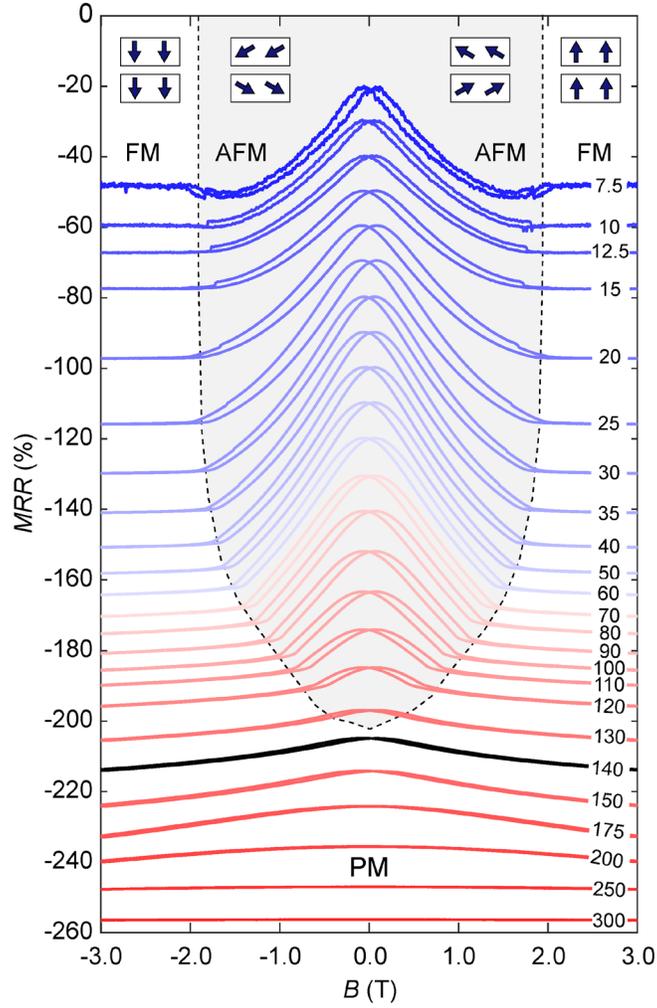

**Figure 14: Magnetotransport properties of 4 L CrSBr.** Magnetoresistance ratio defined as $MRR(B) = \frac{R(B)-R(B=0)}{R(B=0)} * 100$ versus magnetic field at various temperatures with the field oriented along the *c*-axis. Both forward and backward magnetic field scans at each temperature are presented. The curves are offset for clarity. The solid black line is the curve taken at a temperature near $T_N$. The AFM, FM, and PM phases are labeled, and the phase boundary is denoted by a dashed black line. Schematics showing the orientation of the spins in the AFM and FM state are given in the inset.

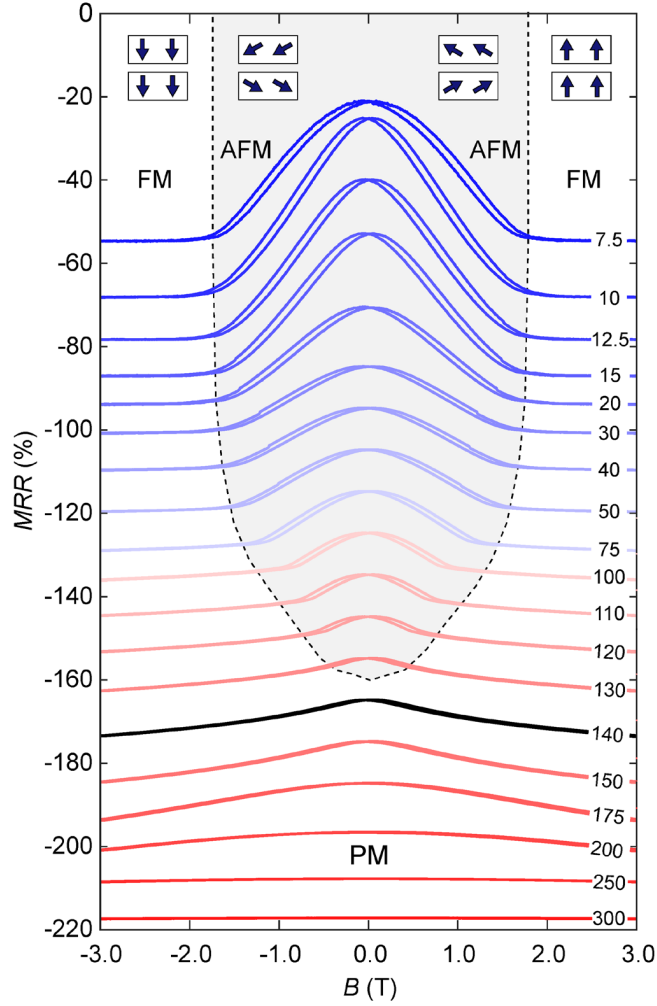

**Figure 15: Magnetotransport properties of 3 L CrSBr.** Magnetoresistance ratio defined as $MRR(B) = \frac{R(B)-R(B=0)}{R(B=0)} * 100$ versus magnetic field at various temperatures with the field oriented along the *c*-axis. Both forward and backward magnetic field scans at each temperature are presented. The curves are offset for clarity. The solid black line is the curve taken at a temperature near $T_N$. The AFM, FM, and PM phases are labeled, and the phase boundary is denoted by a dashed black line. Schematics showing the orientation of the spins in the AFM and FM state are given in the inset.

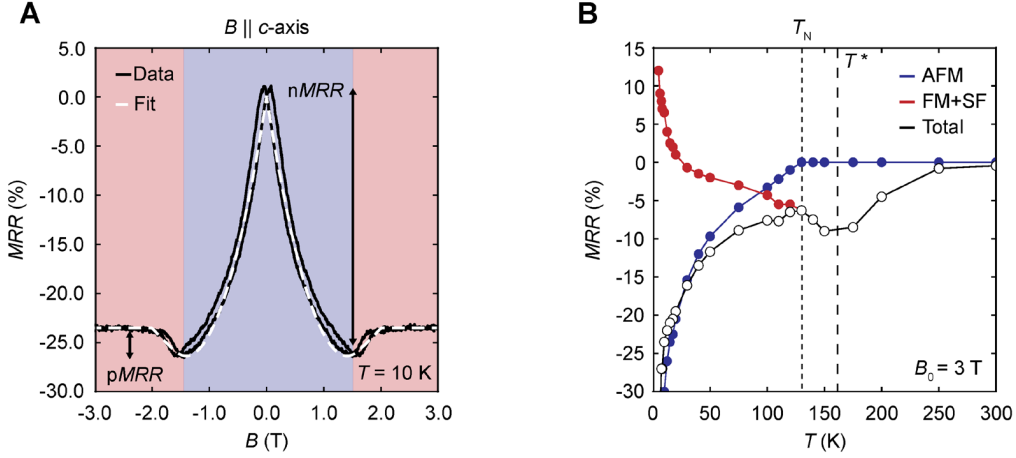

**Figure 16: p*MRR* and n*MRR* components extracted from bilayer CrSBr *MRR*. A)** *MRR* versus field at 10 K for the bilayer sample presented in the main text. The regions of n*MRR* from the layered AFM ordering and the high-field p*MRR* are denoted by blue and red boxes, respectively. The dashed white line is a fit of the form $MRR \propto A_{\text{NMRR}}\left[\left(1 - \frac{B}{B_{\text{sat}}^{\text{NMRR}}}\right)^2 - 1\right] + A_{\text{PMRR}}\left(\frac{B}{B_{\text{sat}}^{\text{PMRR}}}\right)^2 + O(B)$, where $A_{\text{NMRR}}$, $B_{\text{sat}}^{\text{NMRR}}$ and $A_{\text{PMRR}}$, $B_{\text{sat}}^{\text{PMRR}}$ are the amplitudes and corresponding saturation fields for the n*MRR* and p*MRR* components, respectively. The $O(B)$ term encompasses the high-field linear component observed for higher temperatures. **B)** Plot of the n*MRR* component $A_{\text{NMRR}}$ (solid blue dots and line), the p*MRR* component $A_{\text{PMRR}}$ plus the high-field component $O(B)$ (solid red dots and line), and the total *MRR* (solid white dots and black line) versus temperature. The magnetic transition temperatures $T_N$ and $T^*$ are labelled by dashed black lines. The extracted p*MRR* plus high-field component qualitatively reproduces the behavior of monolayer CrSBr, suggesting the p*MRR* in few-layer and bulk CrSBr is related to intraplanar *MRR* effects.

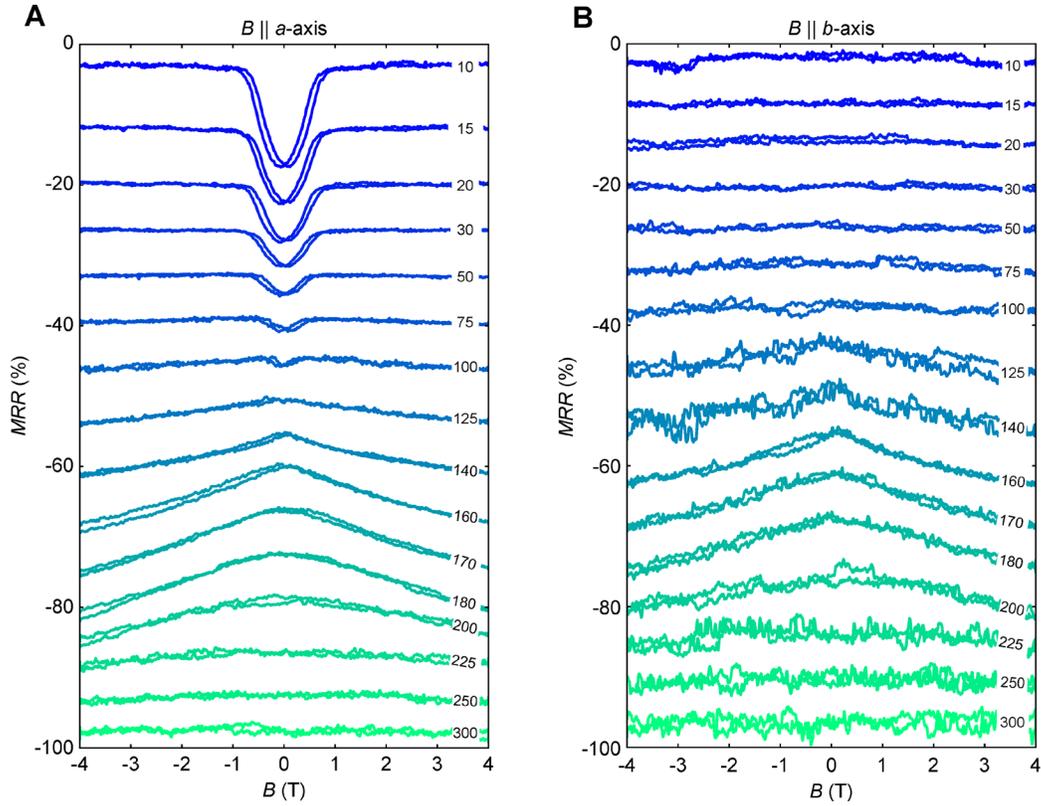

**Figure 17: Additional magnetoresistance measurements of monolayer CrSBr.** *MRR* versus magnetic field at various temperatures with the field oriented along the *a*-axis (**A**) and *b*-axis (**B**) for the monolayer CrSBr device presented in the main text. Both forward and backward magnetic field scans at each temperature are presented. The curves are offset for clarity.

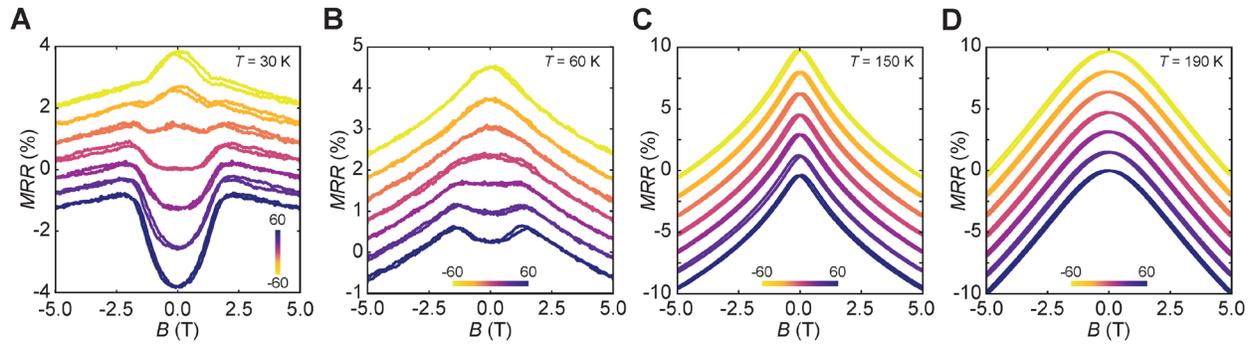

**Figure 18: Gate dependence of monolayer CrSBr *MRR* at higher temperatures. A-D)** *MRR* versus magnetic field as function of back-gate voltage at $T = 30$ K (**A**), $T = 60$ K (**B**), $T = 150$ K (**C**), and $T = 190$ K (**D**) for the monolayer device presented in the main text. In each plot, both forward and backward magnetic field scans are presented. The curves are offset for clarity. The corresponding range of back-gate values is given in the inset of the plots.

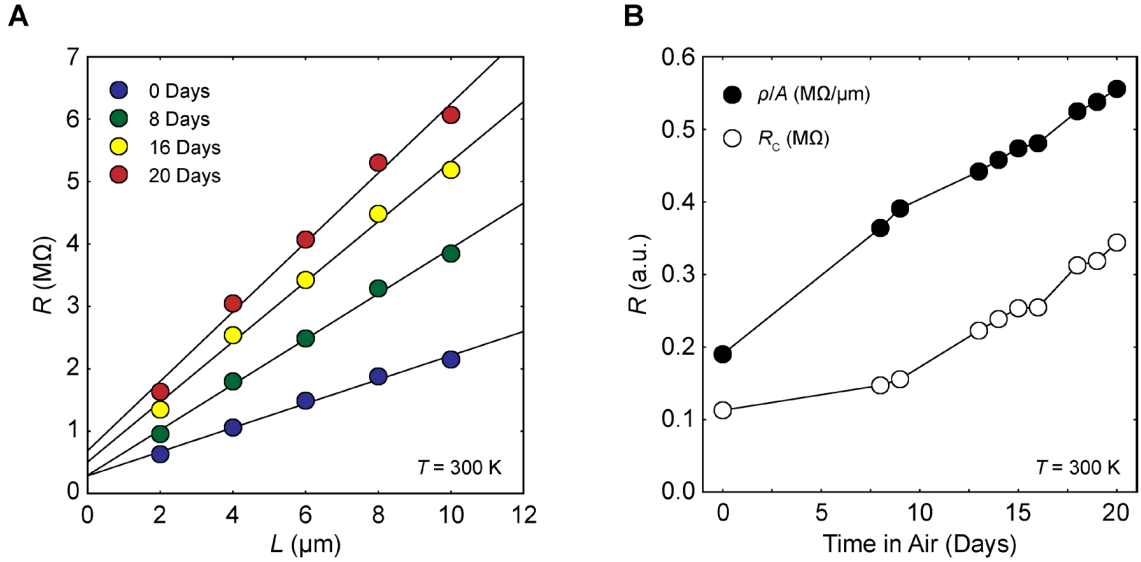

**Figure 19: Oxidation dependence of monolayer CrSBr sample and contact resistances. A)** Transmission line measurement (TLM) of the monolayer CrSBr sample presented in the main text after removing it from the fridge (solid blue dots) and after sitting under ambient conditions for 8 days (solid green dots), 16 days (solid yellow dots), and 20 days (solid red dots). Linear fits to the data are given by solid black lines. **B)** Extracted sample resistivity per cross-sectional area (solid black dots) and contact resistance (solid white dots) versus time spent under ambient conditions. All measurements were taken at room temperature.

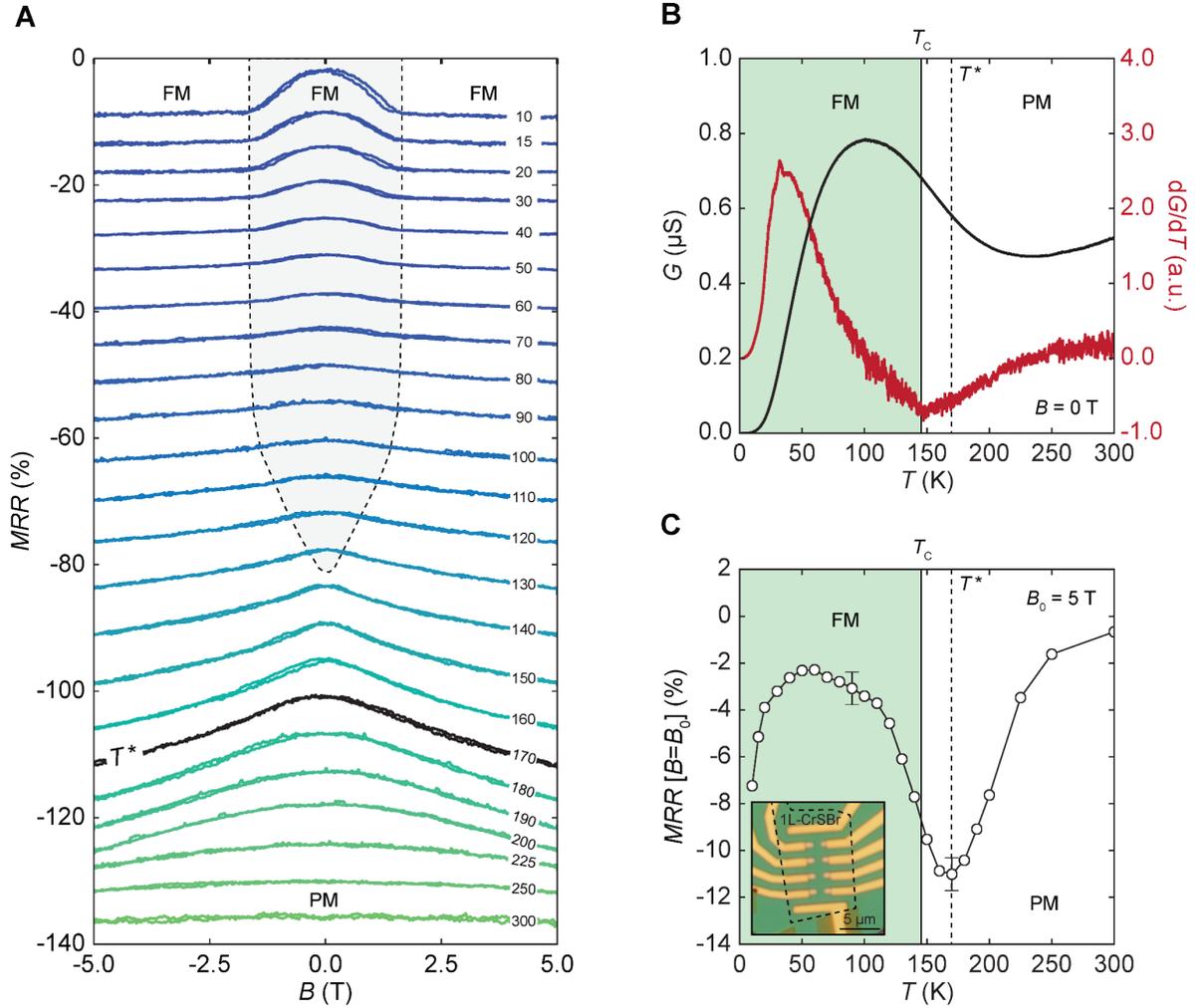

**Figure 20: Transport properties of a second monolayer CrSBr device. A)** Magnetoresistance ratio $MRR(B) = \frac{R(B)-R(B=0)}{R(B=0)} * 100$ versus magnetic field at various temperatures with the field oriented along the *c*-axis. Both forward and backward magnetic field scans at each temperature are presented. The curves are offset for clarity. The solid black line is the curve taken at a temperature near *T**. The FM and PM phases are labeled, and the phase boundary is denoted by a dashed black line. **B)** Conductance (solid black line) and derivative of conductance (solid red line) versus temperature at zero magnetic field. The FM and PM phases are denoted by green and white regions, respectively. **C)** *MRR* at a fixed magnetic field versus temperature. The magnetic field at which the fixed-field *MRR* is calculated is $B_0 = 5$ T. A clear minimum in the fixed-field *MRR* versus temperature at *T** is demarcated by a dashed black line. The FM and PM phases are denoted by green and white regions, respectively. An optical image of the monolayer device is given in the inset. The CrSBr flake is outlined by a dashed black line.

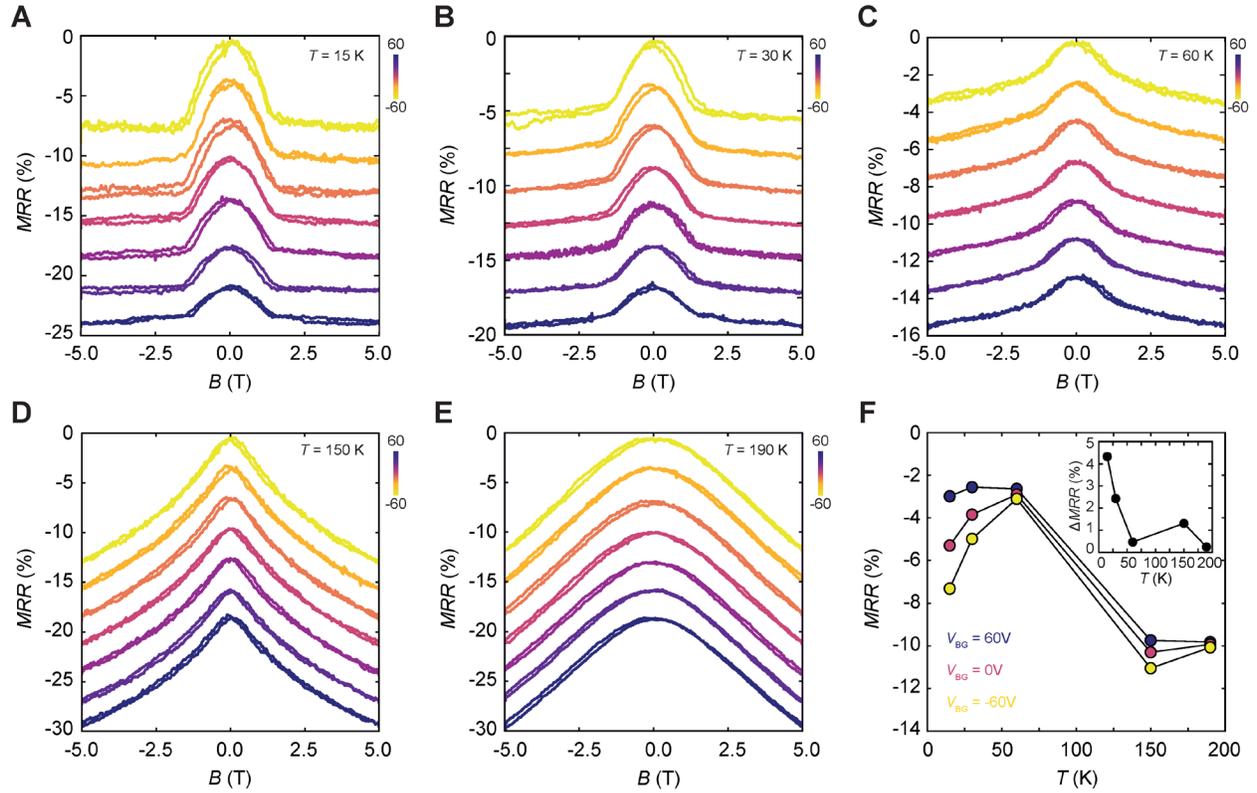

**Figure 21: Gate dependence of *MRR* for a second monolayer CrSBr device. A-E)** *MRR* versus magnetic field for various back-gate voltages at $T = 15$ K (**A**), $T = 30$ K (**B**), $T = 60$ K (**C**), $T = 150$ K (**D**), and $T = 190$ K (**E**) for a second monolayer CrSBr device. In each plot, both forward and backward magnetic field scans are given. The corresponding back-gate voltage ranges are given on the side of each plot. **F)** Fixed-field *MRR* versus temperature for back-gate voltages of $V_{BG} = 60$ V (blue dots), $V_{BG} = 0$ V (pink dots), and $V_{BG} = -60$ V (yellow dots). The total change in fixed-field *MRR* versus back-gate voltage at each temperature is given in the inset. The fixed-field *MRR* was calculated at $B_0 = 5$ T.

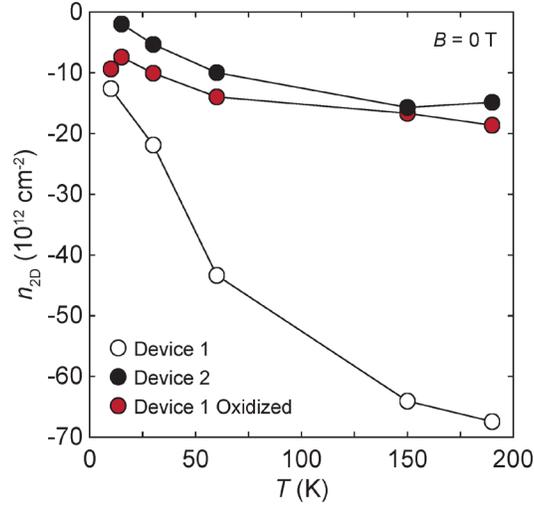

**Figure 22: Estimation of electronic carrier density versus temperature for monolayer CrSBr devices.** Calculated electronic carrier density versus temperature for device 1 (solid white circles), device 2 (solid black circles), and device 1 after oxidizing (solid red circles). The carrier density was estimated by fitting the conductance versus back-gate voltage to a linear dependence and extrapolating the intrinsic carrier density (see Supplemental text for details).

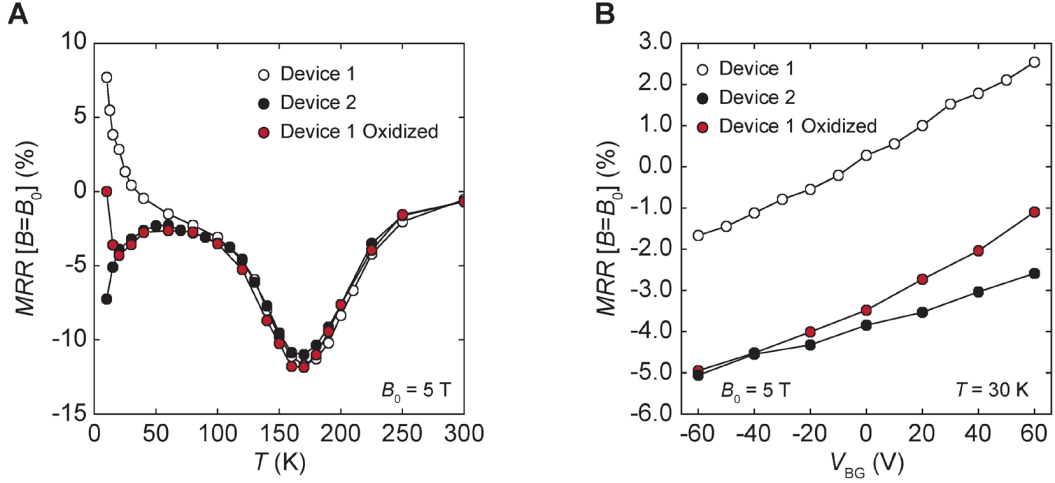

**Figure 23: Comparison of *MRR* between monolayer CrSBr devices. A)** Fixed-field *MRR* versus temperature for device 1 (solid white dots), device 2 (solid black dots), and device 1 after oxidizing (solid red dots). **B)** Fixed-field *MRR* versus back-gate voltage at 30 K for device 1 (solid white dots), device 2 (solid black dots), and device 1 after oxidizing (solid red dots). For both (**A**) and (**B**), the magnetic field at which the fixed-field *MRR* is determined is $B_0 = 5$ T.

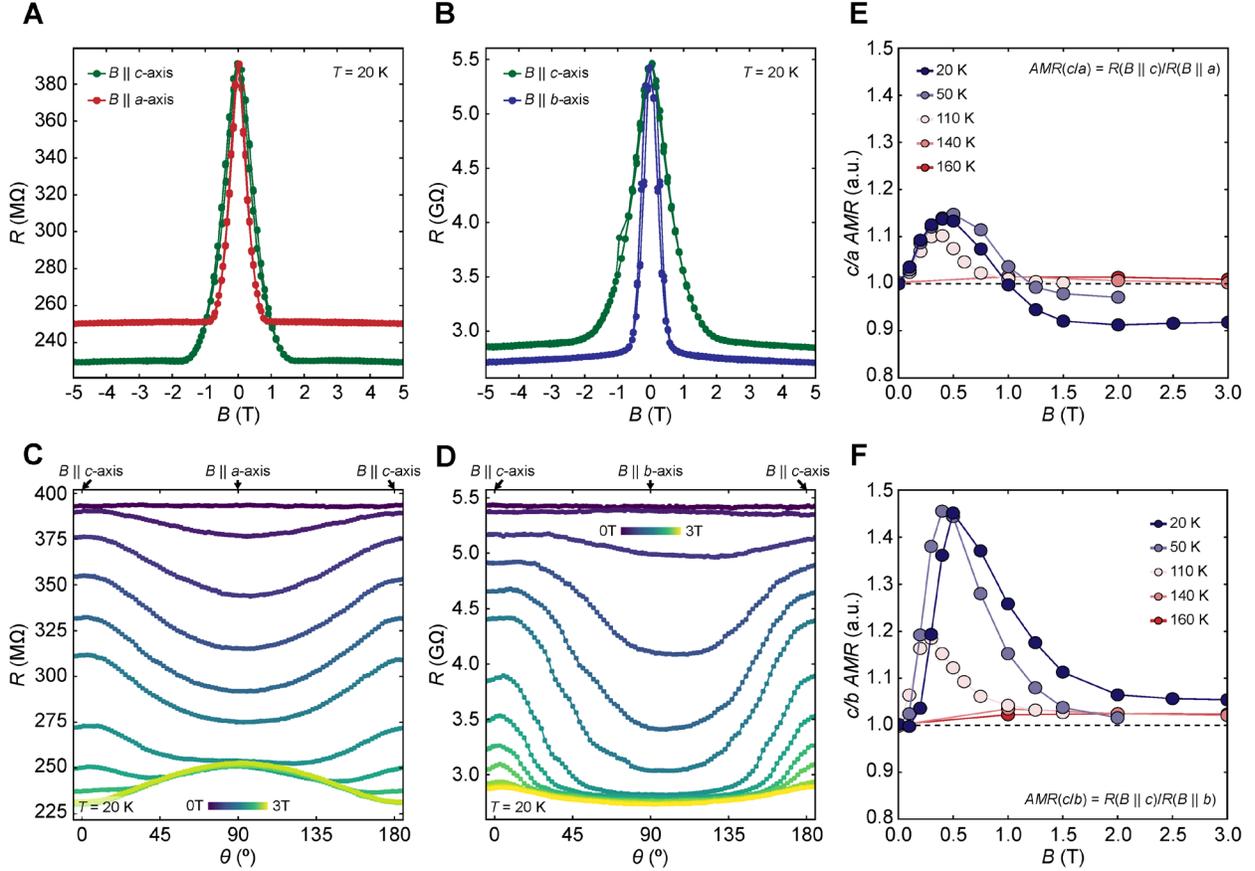

**Figure 24: Field-direction dependence of bulk CrSBr *MRR*. A)** Resistance versus magnetic field for fields parallel to the *c*-axis (green dots) and *a*-axis (red dots) at 20 K measured in a single bulk CrSBr device. **B)** Resistance versus magnetic field for fields parallel to the *c*-axis (green dots) and *a*-axis (blue dots) at 20 K measured in a second bulk CrSBr device. For (**A**) and (**B**), both devices were fabricated from the same single CrSBr crystal. **C, D)** Resistance versus rotator axis angle for various applied magnetic fields for the device in **A** (**C**) and **B** (**D**) at 20 K. The corresponding crystal axis orientations are given above each plot. The color of each trace corresponds to a different magnetic field value, the range of which is denoted in the inset. **E, F)** Extracted anisotropy versus magnetic field for various temperatures comparing the *c*- and *a*-axes (**E**) and the *c*- and *b*-axes (**F**). Anisotropy is defined as $AMR = R(B \parallel x)/R(B \parallel y)$, where x and y are crystal axes. Above the saturation field at 20 K, $R(B \parallel a) > R(B \parallel c) > R(B \parallel b)$, consistent with measurements on monolayer CrSBr.